%% file: MAM_discretization_arxiv-v1.tex
\documentclass{article} 

\usepackage[utf8]{inputenc} 
\usepackage{amsmath}
\usepackage{bbm}
\usepackage{amsfonts}
\usepackage{amsthm}
\usepackage{epsfig}
\usepackage{graphicx}
\usepackage{float}
\usepackage{epstopdf}
\usepackage{pdflscape}
\usepackage[round]{natbib}
\usepackage{bbm}
\usepackage{booktabs}
\usepackage{pgfplots}

\theoremstyle{definition}
\newtheorem{remark}{Remark}

\usepackage{geometry} 
\usepackage{graphicx} 

\usepackage{booktabs} 
\usepackage{array} 
\usepackage{paralist} 
\usepackage{verbatim} 
\usepackage{subfig} 

\usepackage{fancyhdr} 
\usepackage{sectsty}
\allsectionsfont{\sffamily\mdseries\upshape} 

\usepackage[nottoc,notlof,notlot]{tocbibind} 
\usepackage[titles,subfigure]{tocloft} 

\title{Discretization of Lévy semistationary processes with application to estimation}
\author{Mikkel Bennedsen\thanks{
(Corresponding author) CREATES and Department of Economics and Business, 
Aarhus University, 
Fuglesangs All\'e 4,
8210 Aarhus V, Denmark,
E-mail:\ 
\texttt{mbennedsen@econ.au.dk}.
}, 
Asger Lunde\thanks{
CREATES and Department of Economics and Business, 
Aarhus University, 
Fuglesangs All\'e 4,
8210 Aarhus V, Denmark,
E-mail:\ 
\texttt{alunde@econ.au.dk}.
} 
and Mikko S. Pakkanen\thanks{
CREATES and Department of Economics and Business, 
Aarhus University, 
Fuglesangs All\'e 4,
8210 Aarhus V, Denmark,
E-mail:\ 
\texttt{mpakkanen@econ.au.dk}.
}}

\begin{document}\sloppy
\maketitle

\begin{abstract}

Motivated by the construction of the It\^{o} stochastic integral, we consider a step function method to discretize and simulate volatility modulated Lévy semistationary processes. Moreover, we assess the accuracy of the method with a particular focus on integrating kernels with a singularity at the origin. Using the simulation method, we study the finite sample properties of some recently developed estimators of realized volatility and associated parametric estimators for Brownian semistationary processes. Although the theoretical properties of these estimators have been established under high frequency asymptotics, it turns out that the estimators perform well also in a low frequency setting.

\end{abstract}


\smallskip

\begin{msc} 65C05, 62M07 (primary), 62C07 (secondary)  \end{msc}

\section{Introduction}

\citet{ambitTC07} have recently introduced a general and flexible class of tempo-spatial random fields called \emph{ambit fields}. These random fields have been applied in various areas, including modeling of tumour growth (\cite{tumour07}, \cite{ambitTC07}), turbulence (\cite{ole_turbulence03}, \cite{ole-jurgen09}) and finance (\cite{ole_fred_almut11}, \cite{ole_fred_almut13}). For a general reference on the ambit stochastics framework, we refer to \cite{ole_feb_almut12}.

In particular, attention has been given to a class of null-spatial ambit fields, \emph{Lévy semistationary} ($\mathcal{LSS}$) processes and their subclass of \emph{Brownian semistationary}  ($\mathcal{BSS}$)  processes. While these processes are typically neither Markovian nor semimartingales, they are naturally applicable to a wide range of fields including physics, biology and finance. An $\mathcal{LSS}$ process is defined via a stochastic integral of a deterministic kernel function with respect to a driving Lévy process that is subject to volatility modulation. $\mathcal{LSS}$ models provide a flexible, parsimonious and analytically tractable framework, which extends several well-known models, such as the Ornstein-Uhlenbeck (OU) model, continuous time autoregressive-moving-average (CARMA) processes, fractional Brownian motion and more, see e.g.\ \cite{ole_fred_almut13}. In addition, the $\mathcal{LSS}$ framework allows one to go beyond these familiar models and consider processes exhibiting non-standard features such as non-Markovianity, non-semimartingality and long-range dependence. Recently, $\mathcal{LSS}$ models have succesfully been used in the modeling of electricity prices \citep{veraart14} whereas the sub-class of Brownian semistationary processes --- i.e.\ $\mathcal{LSS}$ processes driven by Brownian motion --- have been used in the study of turbulence \citep{ole-jurgen09} and of energy markets \citep{bennedsen14}. The generality and flexibility of the model together with promising early applications has prompted an increasing amount of interest and research in the theoretical properties of the model. 

The strong correlations exhibited by the increments of a typical $\mathcal{LSS}$ process cause the standard estimators of realized volatility introduced in the semimartingale framework to be inadequate. For this reason, \cite{ole_jose_mark09}, \cite{ole_jose_mark11} and \cite{ole_jose_mark13} developed a theory of multipower variations (MPV) for $\mathcal{BSS}$ processes which allowed \cite{ole_mikko_jurgen13} to derive estimators of integrated volatility (IV) and of realized relative volatility (RRV), while parametric estimation in the model --- in particular the estimation of the \emph{smoothness parameter} --- was developed by \cite{ole_jose_mark11} and \cite{jose_emil_mikko_mark13}. These theoretical advances provide an important step towards applying $\mathcal{BSS}$-based models in practice. The theoretical underpinning of these estimators, however, relies on (high frequency) infill asymptotics, that is, on the assumption that the number of observations in a given interval approaches infinity. Naturally, this raises a question concerning the finite sample performance of these estimators --- particularly relevant in applications where the number of observations can be relatively low, such as in (some areas of) finance and particularly in energy markets, where spot prices are observed daily. For this reason, we explore in this paper the finite sample properties of the aforementioned estimators in a low-frequency setting.

The contribution of this paper is twofold. First we present a thorough analysis of the natural method of simulating volatility modulated Lévy semistationary via discretizations inspired by the definition of the stochastic It\^{o} integral. We highlight important features and pitfalls of the method with an emphasis on $\mathcal{LSS}$ processes constructed using an integrating kernel with a singularity at the origin. Such processes are not semimartingales which affects the simulations significantly. To control the error that arises from the simulation scheme, we derive general estimates for the mean squared error of the simulated path and apply this to assess the error of our main $\mathcal{LSS}$ example, where the integrating kernel is the so-called gamma kernel. Second, we analyze the finite-sample performance of various estimators and test statistics for $\mathcal{BSS}$ processes based on power variations through a Monte Carlo study. In particular, we find that these methods perform well even with relatively few observations.

The paper is structured as follows. Section \ref{sec:model} introduces the $\mathcal{LSS}$ process and its key properties, while Section \ref{sec:simulations} outlines a simple discretization and simulation scheme based on the step function approximation of the stochastic integral. Due to the process (possibly) being non-Markovian and a non-semimartingale, simulation can be time consuming and prone to error and we give examples of and recommendations for efficient and accurate simulation. Section \ref{sec:ass} reviews the theory of power variations for $\mathcal{BSS}$ processes and the associated estimators of the smoothness parameter and of integrated volatility before presenting the finite sample properties of these estimators. Section \ref{sec:concl} concludes.

\section{Lévy semistationary processes}\label{sec:model}

We consider a filtered probability space $(\Omega,\mathcal{F},\{\mathcal{F}_t\},\mathbb{P}),$ satisfying the usual conditions of completeness and right continuity of the filtration, and a stochastic process $Y = \{Y(t)\}_{t \in \mathbb{R}_+}$ defined on this space by
\begin{align}\label{eq:LSS}
Y(t) = \mu &+ \int_{\mathbb{R}} h(t-s) a(s) ds + \int_{\mathbb{R}} g(t-s) \sigma(s-) dL(s), 
\end{align}
where $\mu \in \mathbb{R}$ is a constant, $h \in L^1(\mathbb{R})$ and  $g \in L^2(\mathbb{R})$ are deterministic kernel functions such that $h(x)=g(x)=0$ for $x\leq 0.$ $a = \{a(t)\}_{t\in \mathbb{R}}$ and $\sigma = \{\sigma(t)\}_{t\in \mathbb{R}}$ are stochastic processes adapted to the filtration $\{\mathcal{F}_t\}_{t\in \mathbb{R}}$ such that the integrals in \eqref{eq:LSS} exists. We take $L$ to be a two-sided Lévy process  on $\mathbb{R}$ --- that is, we take a Lévy process $L'$ defined on $\mathbb{R}_+$ and an independent copy of it, $L'',$ and define $L(t) = L'(t)$ for $t \geq 0$ and $L(t) = -L''(-(t-))$ for $t < 0.$ The process $Y = \{Y_t\}_{t \in \mathbb{R}}$ in \eqref{eq:LSS} is called a Lévy semistationary $(\mathcal{LSS})$ process; the name being derived from the fact that under suitable conditions, such as $(a,\sigma)$ being stationary and independent of $L,$ the resulting process $Y$ will be (strictly) stationary. This is also the reason for the moving average type kernel and for starting the integration at minus infinity instead of at zero. 

Stationarity is a desirable feature in a range of applications such as turbulence and commodity markets and \eqref{eq:LSS} thus allows us to specify the model \emph{directly in stationarity} as opposed to only achieving stationarity in the limit as $t \rightarrow \infty$ which is the case for some other models, such as the OU process starting at a point $X(0)=x_0 \in  \mathbb{R}.$ The first integral in \eqref{eq:LSS} is a Lebesgue integral and will pose no problems from a simulation standpoint and we will therefore only focus on the part coming from the second integral, that is, from now on we consider the driftless $\mathcal{LSS}$ process
\begin{align}
X(t) &= \int_{\mathbb{R}} g(t-s) \sigma(s-) dL(s) = \int_{-\infty}^{t} g(t-s) \sigma(s-) dL(s). \label{eq:LSS2}
\end{align}

\subsection{Autocorrelation structure}\label{sec:ACF}
In the following we will make extensive use of the flexible autocorrelation structure that the $\mathcal{LSS}$ model \eqref{eq:LSS2} provide. Assume for simplicity that $L$ has mean zero, is square integrable and that $\sigma$ is stationary and independent of $L.$ Now $\mathbb{E}[X(t)]=0$ for all $t$ and, denoting $\kappa = Var(L(1)),$ we have  for $h\geq0$ the covariance function
\begin{align}
\gamma(h) = \mathbb{E}[X(t) X(t+h)] &= \mathbb{E}\left[ \kappa \int_{-\infty}^t g(t-s)g(t+h-s)\sigma^2(s) ds \right] \nonumber \\
&=  \kappa \mathbb{E}[\sigma^2(0)]  \int_0^{\infty} g(x)g(x+h)dx, \label{eq:ACVF}
\end{align}
from which we see that the kernel function $g$ gives us control over the correlation function of the process. This allows us to capture, in a flexible way, a wide range of correlation structures inspired e.g.\ by theoretical or empirical considerations. An example of this is given in Section \ref{sec:ass_sim} where we show that for a particular choice of kernel function $g,$ \eqref{eq:ACVF} gives rise to the well-known Mat\'ern  covariance function \citep{matern60} which is used in a variety of fields such as in machine learning and in the study of turbulence.

\section{Simulation of Lévy semistationary processes}\label{sec:simulations}

Consider the problem of simulating $N+1 \in \mathbb{N}$ points of the $\mathcal{LSS}$ process on an equidistant grid with step size $\delta,$ $X(i\delta), i =0, 1 , \ldots N.$  The general simulation problem involves truncation and approximation of the integral and will be covered in Section \ref{sec:generalsim} below, but consider first the (important) case of the $\mathcal{BSS}$ process without stochastic volatility, i.e.\ where $L=W$ is a Brownian motion and volatility is constant, $\sigma(t) = \sigma \in \mathbb{R}_+$ for all $t.$ Now, the process $X  = \{X(t)\}_{t \in \mathbb{R}}$ in \eqref{eq:LSS2} is a mean zero Gaussian process with covariance given by \eqref{eq:ACVF}. Denoting by $\Sigma = \{\gamma((i-j)\delta)\}_{i,j=0}^N$ the Toeplitz matrix arising from this covariance function we can obtain \emph{exact} simulations of  $X$ by drawing a $(N$$+$$1)$-dimensional standard normal random vector ${\bf Z} \sim N_{N+1}(0,I)$ and setting ${\bf X} := (X_0,X_1,\ldots,X_N)^T = F' {\bf Z}$ where $F$ is the Cholesky decomposition of $\Sigma.$

\subsection{Discretizing the $\mathcal{LSS}$ process} \label{sec:generalsim}

Although there do exist alternative schemes for simulating general $\mathcal{LSS}$ processes (see \cite{heidar13} and \cite{heidar13fourier}) we consider here the simpler route of a step function approximation of \eqref{eq:LSS2}, which was also done by, e.g., \cite{emil13}. To motivate this approach, write for $i=0, 1, \ldots N,$
\begin{align*}
X(i\delta) &= \int_{\mathbb{R}} g(i\delta-s) \sigma(s-) dL(s) = \sum_{j=-\infty}^{\infty} \int_{(j-1)\delta}^{j\delta} g(i\delta-s) \sigma(s-) dL(s).
\end{align*}
Now, if both $g$ and $\sigma$ are approximately constant and equal to the left end point value on the intervals $[(j-1)\delta,j\delta),$ we get
\begin{align}\label{eq:approx}
X(i\delta)  &\approx \sum_{j=-\infty}^{\infty} g((i-j+1)\delta) \sigma((j-1)\delta)  \Delta L_{j},
\end{align} 
where $ \Delta L_{j} := L(\delta j) - L(\delta(j-1)),$ $j \in \mathbb{N},$ are the increments of the background driving L\'{e}vy process. Note that this approximation is quite natural as it is similar to the one used through simple (step) functions in the construction of the It\^{o} integral in \eqref{eq:LSS2}. Denoting $G(i-j):=g((i-j+1)\delta)$ and $\Sigma(j):=\sigma((j-1)\delta)  \Delta L_{j}$ we now have
\begin{align}\label{eq:conv}
X(i\delta)  &\approx  \left(G \ast \Sigma \right) (i),
\end{align} 
where $\ast$ denotes (discrete) convolution. In other words, given that the step function approximation is reasonable, the $\mathcal{LSS}$ process is approximately a discrete convolution of the kernel function $g$ with the stochastic part consisting of the volatility process and the increments of the driving Lévy process, $\sigma \Delta L.$ The upshot of the approximation \eqref{eq:conv} is that fast simulation becomes possible since most software packages come with extremely fast and efficient numerical methods for computing convolutions.\footnote{We used the MATLAB function \texttt{fftconv} available from the MATLAB Central which proved to be far superior in terms of a speed/accuracy trade-off as compared to the built-in MATLAB function \texttt{conv}.} Further, these algorithms will not only output the simulated value of the process at a time point $t = i\delta,$ but will output the whole vector of desired values $X(j\delta),$ $j = 0, 1, \ldots, N.$ Appendix \ref{sec:app_sim} contains a step-by-step routine for the simulations performed in this paper. Note, that in practice when simulating, we need to truncate the infinite sum \eqref{eq:approx} from below at a level $-M\delta$ for some $M \in \mathbb{N},$ see also the next section.

\subsection{Controlling simulation error by subsampling}\label{sec:errorALT}

\begin{figure*}[!t] 
\centering 
\includegraphics[scale=.85]{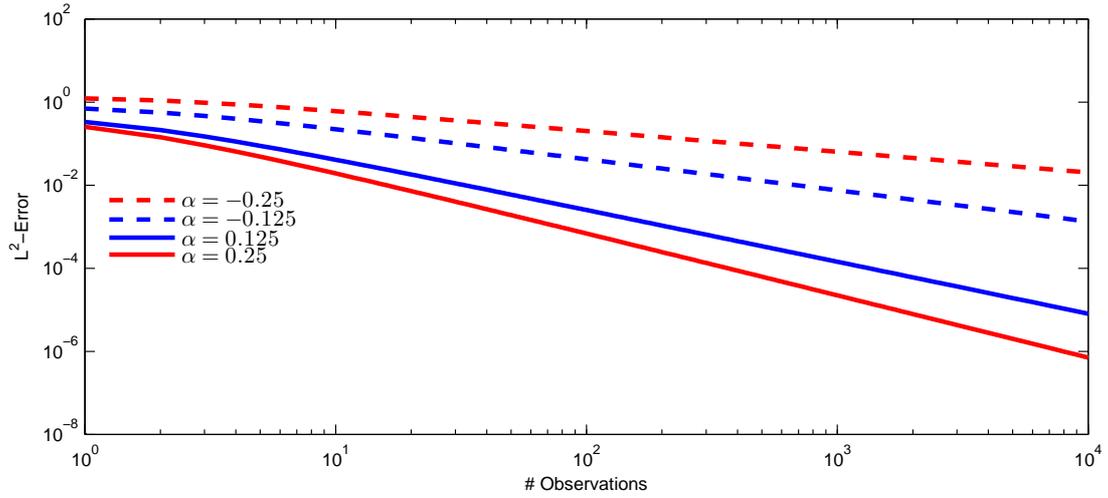} 
\caption{\it Simulation error introduced by truncation and step function approximation of a $\mathcal{LSS}$ process with $\mathbb{E}[\sigma^2(0)]=\kappa=1$ and $\lambda=1.$ We see how the simulation error is significantly larger for negative values of $\alpha$ (dashed lines) as compared to the positive values of $\alpha$ (solid lines) caused by the the singularity at the origin in the kernel of the former. Note the log-log-scale.}
\label{fig:fullErr}
\end{figure*}

Unless we are in the Gaussian case of no stochastic volatility, we will naturally introduce error when simulating; both from the step function approximation as well as from the truncation of the sum at $j=-M.$ It is possible to derive some bounds on the errors introduced, which we consider in the following. Let $\tilde{X}(t) = \int_{\mathbb{R}} \tilde{g}(t-s)\tilde{\sigma}(s-) dL(s)$ be an approximation of $X(t),$ where we have altered the kernel function $g$ and the stochastic volatility component $\sigma$ in such a way as to make simulation of $\tilde{X}$ feasible. Now, for a given time $t=i\delta,$ the $L^2$-error of the approximation is given by
\begin{align*}
\| X(t) - \tilde{X}(t) \|_2^2 &:= \mathbb{E}[|X(t)-\tilde{X}(t)|^2] = \mathbb{E}\bigg[|\int_{\mathbb{R}}(g(i\delta-s)\sigma(s-) - \tilde{g}(i\delta-s)\tilde{\sigma}(s-) )dL(s) |^2\bigg] \\
&:= C_1 + C_2 + C_3,
\end{align*}
where the constants $C_i,$ $i =1, 2, 3$ are given by simply expanding the square and applying the stochastic Fubini theorem,
\begin{align*}
C_1 &= \kappa \int_{\mathbb{R}} g^2(i\delta-s) \mathbb{E}[\sigma^2(s)] ds \\
C_2 &= \kappa \int_{\mathbb{R}} \tilde{g}^2(i\delta-s) \mathbb{E}[\tilde{\sigma}^2(s)] ds \\
C_3 &=  -2\kappa \int_{\mathbb{R}} g(i\delta-s) \tilde{g}(i\delta-s) \mathbb{E}[\sigma(s)\tilde{\sigma}(s)] ds.
\end{align*}
To recover the truncated step function approximation introduced in the previous section we let $\tilde{g}(x) = g(j\delta) \mathbbm{1}_{(0,t+M\delta]}(x)$ and $\tilde{\sigma}(x) = \sigma(\lfloor x/\delta\rfloor\delta),$ where $\lfloor x \rfloor$ denotes the integer part of $x.$ As an illustration, consider our main example, the gamma kernel $g(x) = x^{\alpha} e^{-\lambda}$ with $\alpha > -\tfrac{1}{2}$ and $\lambda >0.$ Note, that for $\alpha <0,$ $g$ has a singularity at the origin and for $\alpha \in (-\tfrac{1}{2},0) \cup (0,\tfrac{1}{2}),$ $X$ will not be a semimartingale. To ease the exposition in deriving the following estimates on the error, we also assume that $\sigma = \{\sigma(t)\}_{t \in \mathbb{R}}$ is a martingale (otherwise, the expression for $C_3$ will be more complicated). In this case we have 
\begin{align*}
C_1 &= \kappa \mathbb{E}[\sigma^2(0)]  (2\lambda)^{-2\alpha-1} \Gamma(2\alpha+1) \\
C_2 &= \kappa\mathbb{E}[\sigma^2(0)] \delta \sum_{j=1}^{i+M} (j\delta)^{2\alpha} e^{-2\lambda j \delta}  \\
C_3 &= -2 \kappa \mathbb{E}[\sigma^2(0)]  \lambda^{-\alpha-1} \sum_{j=1}^{i+M}  (j\delta)^{2\alpha} e^{-2\lambda j \delta}\left(\gamma(\alpha+1,\lambda(j-1)\delta)- \gamma(\alpha+1,\lambda(j-2)\delta) \right),
\end{align*}
where we used equation (3.381.1) in \cite{integrals} and $\Gamma$ is the gamma function and $\gamma$ the (lower) incomplete gamma function. In Figure \ref{fig:fullErr} we see an illustration of how the error decreases when we increase the number of simulated points, $N,$ in the interval $[0,1].$ The error estimates are done for $\alpha = -0.25, -0.125$ (dashed lines) and $\alpha = 0.125, 0.25$ (solid lines). We clearly see how the singularity of the integrating kernel at the origin for $\alpha<0$ introduces a much higher $L^2$-error in the simulations. Of course, it is possible to increase the number of simulated points $N$ until a desired level of accuracy is reached. For instance, if one is interested only in a fixed number of observations, say $N_0$ (which is the case in our finite sample investigations below), the solution to this problem is to sample at a very fine grid using a large $N = k N_0$ number of observations for some $k \in \mathbb{N}$ and then subsample from these values to get the desired $\mathcal{LSS}$ path with $N_0$ observations. That is, one picks out every $k$-th observation from the simulated path consisting of $N = k N_0$ points.

\subsection{Illustration of simulations}

\begin{figure*}[!t]
\centering
  \includegraphics[scale=0.85]{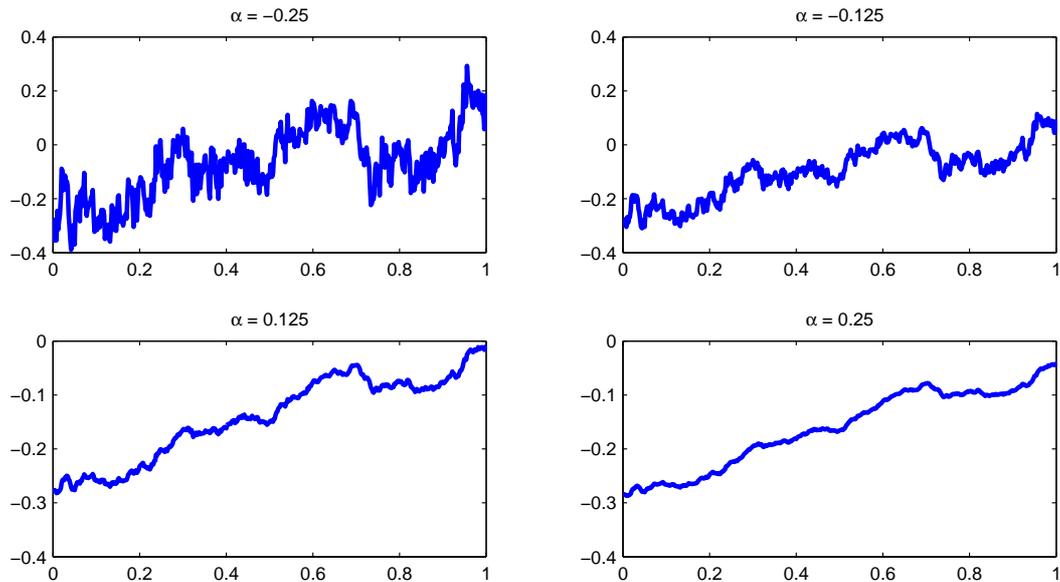}
\caption{\it Simulation of $\mathcal{BSS}$ processes for different values of the smoothness parameter $\alpha$. The kernel function is the gamma kernel $g(x) = x^{\alpha} e^{-\lambda x}$ with $\lambda = 1$ and the number of observations on the interval $[0,1]$ is $N=500.$ The same random numbers have been used, hence the only difference in the four simulations is value of $\alpha$.}
\label{fig:plotBSS}
\end{figure*}

Consider a $\mathcal{LSS}$ process with the gamma kernel, $g(x) = x^{\alpha} e^{-\lambda x}$ for $\alpha > -\tfrac{1}{2}$ and $\lambda >0.$ This kernel function has been shown to be useful in the study of turbulence in e.g.\ \cite{jose_emil_mikko_mark13} and in modelling electricity spot prices in \cite{bennedsen14}. Note, that for $\alpha \in \left( -\tfrac{1}{2},0\right) \cup \left( 0, \tfrac{1}{2} \right),$ the resulting $\mathcal{LSS}$ process $X$ is \emph{neither a semimartingale nor Markovian}. Also, for $\alpha=0$ the process is a L\'{e}vy-driven Ornstein-Uhlenbeck (OU) process (so, our framework generalizes the popular OU models). Note further, that for $\alpha<0$ the kernel has a singularity at $x=0,$ which, as we shall see, significantly encumbers simulation due to the error illustrated in Figure \ref{fig:fullErr}. Figure \ref{fig:plotBSS} shows four simulated $\mathcal{BSS}$ paths using $\lambda=\sigma(t)=1$ for all $t,$ with different values of $\alpha$. It is clear how the value of $\alpha$ controls the smoothness of the process with large negative values of $\alpha$ corresponding to a very rough path while positive values correspond to a smooth path.

\subsection{Assessing the accuracy the simulations}\label{sec:ass_sim}

\begin{figure*}[!t]
\centering
\begin{minipage}{.495\textwidth}
  \centering
  \includegraphics[scale=0.80]{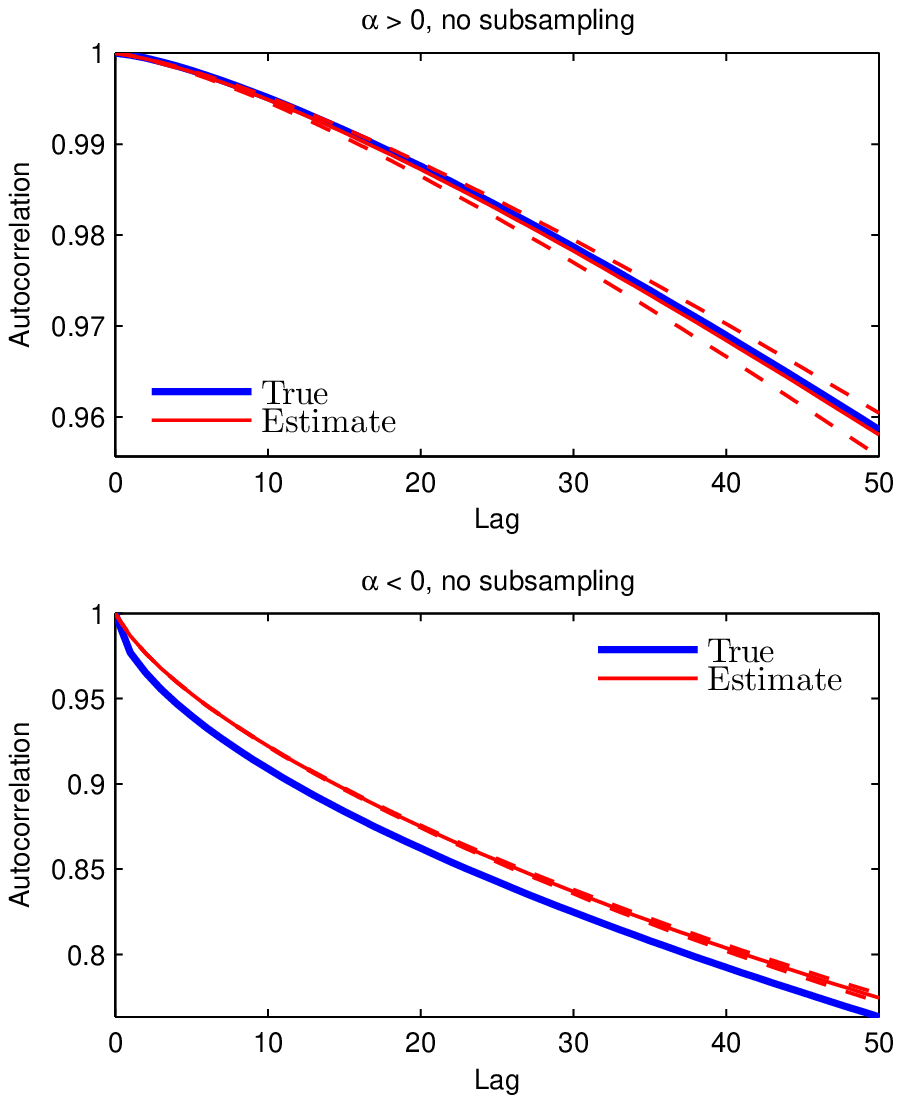}
\end{minipage}
\begin{minipage}{.495\textwidth}
  \centering
   \includegraphics[scale=0.80]{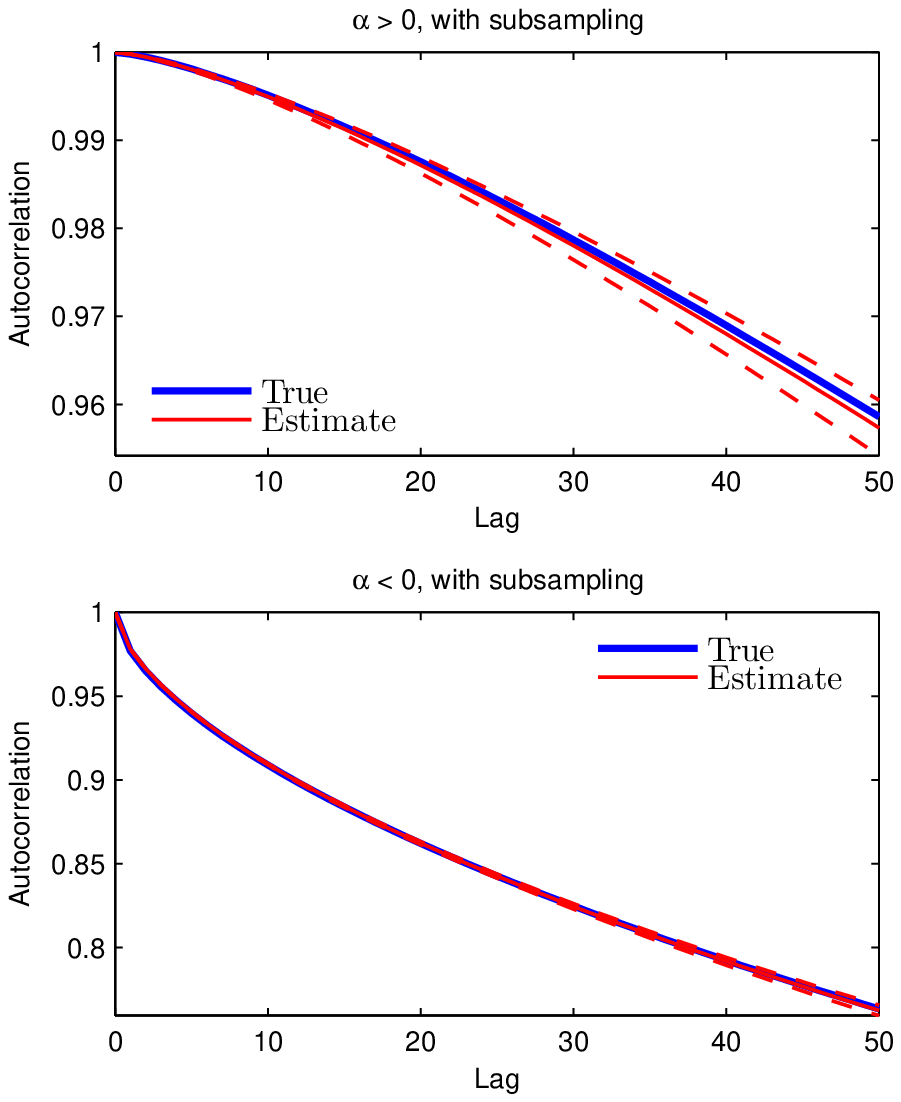}
\end{minipage}
\caption{\it Comparison of the theoretical vs. empirical (simulated) ACF using $\alpha = 0.2$ (top) and $\alpha=-0.2$ (bottom) including $95\%$ numerical confidence bands (dashed). The simulations on the left are without subsampling and those on the right are with subsampling ($k=100$). We see how the error from the simulations when $\alpha<0$ causes the paths to have the wrong second order structure but that this is remedied by sampling at a finer grid.}
\label{fig:ACFcomp}
\end{figure*}

Using the correlation function \eqref{eq:ACVF} derived in Section \ref{sec:ACF} we can check how our simulations perform in terms of how they capture the second order structure of the target theoretical process to be approximated. As an illustration, consider again the $\mathcal{BSS}$ process with a gamma kernel $g(x) = x^{\alpha}e^{-\lambda}.$ Supposing $\{\sigma(t)\}_{t \in \mathbb{R}}$ to be stationary and using equations (3.383.8), (8.331.1) and (8.335.1) in \cite{integrals}, we find the variance of the process to be $Var(X_t) = \gamma(0) = \kappa \mathbb{E}[\sigma^2(0)]\Gamma(2\alpha+1)(2\lambda)^{-(2\alpha+1)}$ and the correlations to be given by the Mat\'{e}rn correlation function (\cite{matern60}, \cite{matern93})
\begin{align*}
 \rho(h) = \frac{\gamma(h)}{\gamma(0)} = \frac{2^{-\alpha+ \frac{1}{2}}}{\Gamma(\alpha+1/2)}(\lambda h)^{\alpha + \frac{1}{2}} K_{\alpha+\frac{1}{2}}(\lambda h),
\end{align*}
where $K_{\nu}(x)$ is the modified Bessel function of the third kind.  Figure \ref{fig:ACFcomp} provides examples of how to assess the accuracy of the simulations and how problems arise when using a kernel with a singularity at the origin: the top plots are with $\alpha = 0.2$ while the bottom plots are with $\alpha = -0.2.$ The two left plots are using the simulation algorithm without subsampling while we in the right plots simulate $k=100$ times as many time points as needed (i.e.\ $kN = 100\cdot 500 = 50,000$) and then subsample the $N=500$ desired time points from this path. The result is an accurate second order structure as can be seen from the bottom right figure. The remaining parameters in the simulation are $\lambda = 1, M=1,000$ and the stochastic volatility is the exponential of a Gaussian Ornstein-Uhlenbeck process, $\log \sigma(t) = \int_{-\infty}^t e^{-\beta (t-s)} dB(s),$ where $\beta=5$ and $B$ is a standard Brownian motion independent of $L.$ 

\section{Application: estimation of $\mathcal{BSS}$ processes}\label{sec:ass}

The following asymptotic results are valid only for $\mathcal{BSS}$ processes, i.e.\ when $L = W$ in \eqref{eq:LSS2} is a Brownian motion and we will from now on work with this process. Analogous research in the general $\mathcal{LSS}$ framework is ongoing. We note that the results also hold for $\mathcal{BSS}$ processes with drift as in \eqref{eq:LSS}, assuming some smoothness conditions of the drift term, see \cite{jose_emil_mikko_mark13}. In Section \ref{sec:al_est} we present the estimator of the smoothness parameter, $\alpha$, developed in \cite{ole_jose_mark11} and in Section \ref{sec:sig_est} we present the estimator of the RRV together with a test for the presence of stochastic volatility in our $\mathcal{BSS}$ process, developed in \cite{ole_mikko_jurgen13}, followed by a study of their finite sample properties.

\subsection{Simulation setup}\label{sec:simset}

In what follows, we assume that $g(x) = L(x)x^{\alpha},$ where $L$ is a slowly varying function at $0,$ i.e.\ $\lim_{x\rightarrow 0} \frac{L(cx)}{L(x)} = 1$ for all $c >0.$ In other words, we require that $g$ behaves as $x^{\alpha}$ at $x=0.$ This condition is obviously fulfilled for our main example, the gamma kernel, which we will use when simulating below. When a stochastic volatility component is present in the simulations, we specify this as the exponential of a Gaussian Ornstein-Uhlenbeck process, $\log \sigma (t) = \int_{-\infty}^t e^{-\beta (t-s)} dB(s),$ where $\beta >0$ and $B$ is a standard Brownian motion possibly correlated with $W.$ This process is Gaussian and Markovian and can thus be simulated in an exact way incurring no simulation error using the recursion
\begin{align*}
\log \sigma(t+\delta) = e^{-\beta \delta} \log \sigma(t) + e^{-\beta \delta}\int_t^{t+\delta} e^{-\beta(t-s)}dB(s) \stackrel{d}{=} e^{-\beta \delta} \log \sigma(t) + \sqrt{\frac{1}{2\beta}(1-e^{-2\beta \delta})}Z,
\end{align*}
where $Z \sim N(0,1).$ See e.g.\ \cite{glasserman} Section 3.3.1.

\subsection{Estimation of the smoothness parameter $\alpha$}\label{sec:al_est}
The parameter $\alpha > -\tfrac{1}{2}$ is called the \emph{smoothness} parameter since it controls the small scale behavior of the paths of the $\mathcal{BSS}$ process $X.$ For $\alpha<0$, $X$ will exhibit very rough paths, while for $\alpha>0$ they will be smooth whereas $\alpha=0$ corresponds to paths of a process driven by a Brownian motion, see Figure \ref{fig:plotBSS} for examples. This behavior of the process for varying the value of $\alpha$ is analogous to the role that the Hurst exponent plays for the fractional Brownian motion (fBm) and the small scale behavior of the $\mathcal{BSS}$ process is actually similar to that of the fBm where the link between the smoothness parameter and the Hurst exponent, $H,$ is given by $H = \alpha + \frac{1}{2} \in (0,1).$ See e.g.\ \cite{Nualart_fbm} for more information about Hurst exponents and their connection to the fractional Brownian motion.

To estimate $\alpha$, let $T>0$ and suppose we have observed the process $X = \{X(t)\}_{t \in \mathbb{R}}$ on an equidistant grid $0=t_0 < t_1 < ...< t_N = T$ with grid size $t_i - t_{i-1} = \delta$ for all $i\geq 1.$ Define the \emph{second order differences at frequency } $v$ to be $\diamondsuit^v_i X := X(i\delta) - 2X((i-v)\delta) + X((i-2v)\delta),$ $i=2v,2v+1,...,N.$ For $p>0$ we also define the associated p'th power variations $V_{v,t}^p := \sum_{i=2v}^{\lfloor t/\delta \rfloor} |\diamondsuit^v_i X|^p.$ 

Now, as proved in  \cite{ole_jose_mark11}, see also \cite{jose_emil_mikko_mark13}, we have the following asymptotic result for the \emph{change of frequency} (COF) estimator $COF(\delta,p) = \frac{V_{1,t}^p}{V_{2,t}^p}$
\begin{align*} 
COF(\delta,p) \stackrel{ucp}{\rightarrow} 2^{\frac{(2\alpha +1)p}{2}},
\end{align*}
as $\delta \rightarrow 0,$  where "ucp" means uniform convergence in probability on compact sets. 
\begin{remark}
This kind of asymptotics is known as infill asymptotics, i.e.\ we consider the time $T>0$ fixed and let the number of observations in $[0,T]$ go to infinity so that the time between successive observations, $\delta,$ goes to zero. 
\end{remark}
Our estimator of the smoothness parameter, $\alpha$, thus becomes
\begin{align} \label{eq:al_lim} 
\hat{\alpha}(\delta,p) = \frac{\log_2 (COF(\delta,p))}{p}-\frac{1}{2}
\end{align}
and we have $\hat{\alpha}(\delta,p) \stackrel{ucp}{\rightarrow} \alpha$ as $\delta \rightarrow 0$ for all $p>0.$ The estimator \eqref{eq:al_lim} is a \emph{feasible} estimator in the sense that it only depends on the observed data, $X(i\delta), i=0,1,...N.$ 

\begin{figure*}[!t]
\centering
\begin{minipage}{.5\textwidth}
  \centering
  \includegraphics[scale=0.85]{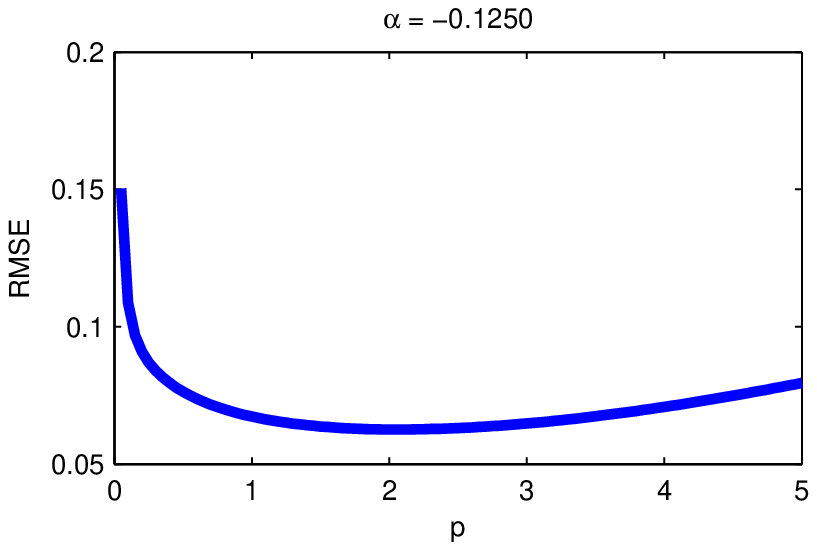}
\end{minipage}%
\begin{minipage}{.5\textwidth}
  \centering
   \includegraphics[scale=0.85]{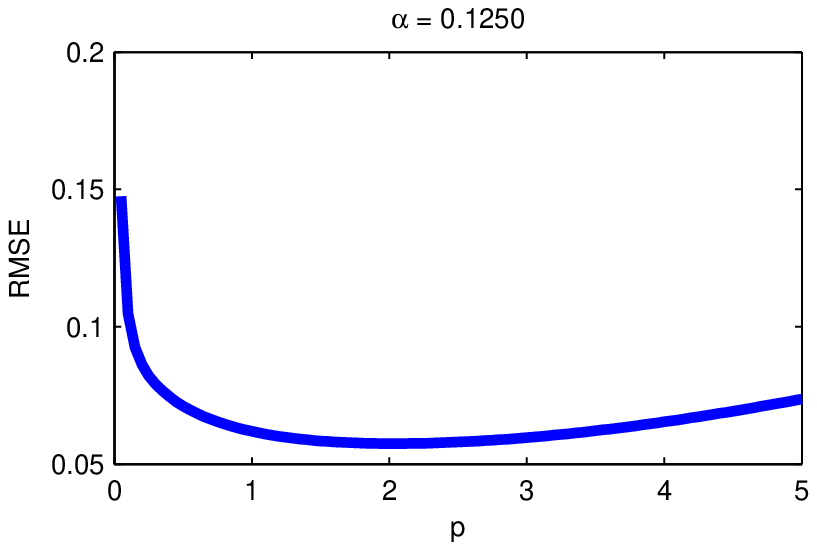}
\end{minipage}
\begin{minipage}{.97\textwidth}
  \centering
  \includegraphics[scale=0.85]{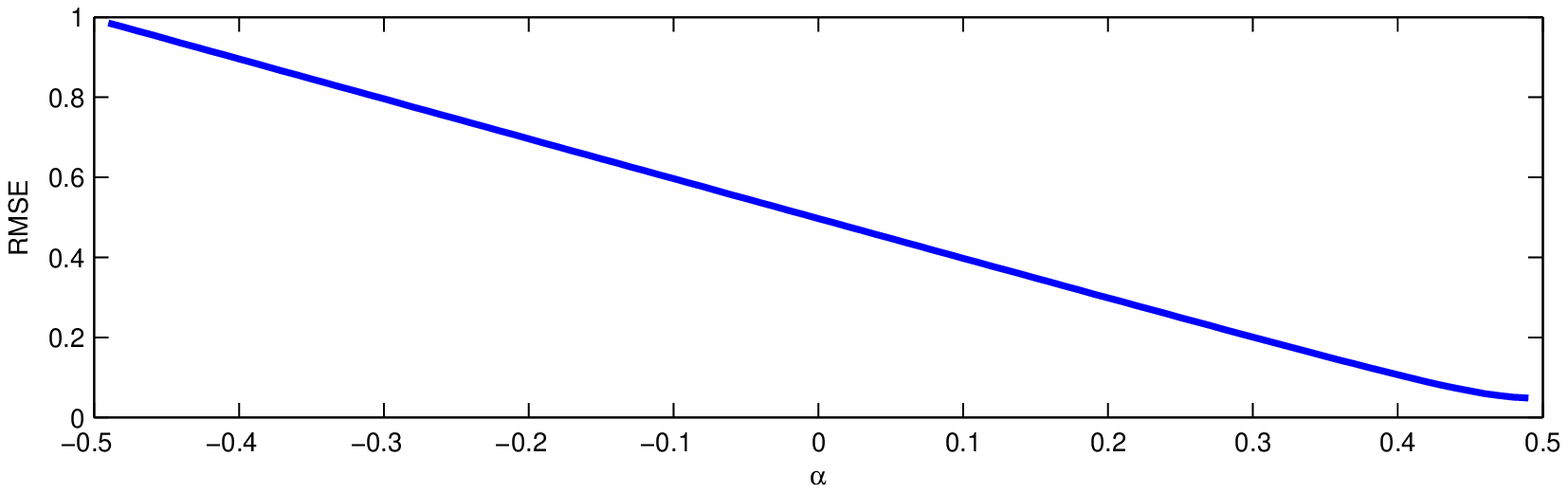}
\end{minipage}
\caption{\it Investigation of the performance of the COF estimator for varying values of the exponent $p$ (top) and various values of $\alpha$ (bottom). The root mean squared error is minimized for $p=2$ for both $\alpha<0$ and $\alpha>0.$ In the bottom plot we see that the RMSE is decreasing as $\alpha$ increases. 20,000 Monte Carlo simulations with $\lambda=1 = \sigma(t) = 1$ for all $t$ and $N=500.$}
\label{fig:pStudy}
\end{figure*}

In addition to the estimator of $\alpha$, \cite{jose_emil_mikko_mark13} provide an associated central limit theorem for $\alpha \in \left(-\frac{1}{2},\frac{1}{4}\right)$:
\begin{align}\label{eq:clt} 
\frac{(\hat{\alpha}(\delta,p)-\alpha)V_{2,t}^p\log(2)p}{\sqrt{m_{2p}^{-1}V_{2,t}^{2p}e^T_1\Lambda_p e_1}} \stackrel{d}{\rightarrow} N(0,1),
\end{align}
where $m_p = E[|U|^p],$  $U \sim N(0,1),$ $e_1 = (-1,1)^T$ and $\Lambda_p = \Lambda_p(\alpha)$ is a 2-by-2 matrix depending on $\alpha$. These results require a choice for the exponent, $p,$ used in the calculations of the power variations, the standard choice being $p=2,$ yielding squared returns in the power variations, and we will also do this here. In Figure \ref{fig:pStudy} we see a justification of this as the root mean squared error of the estimator of $\alpha$ in \eqref{eq:al_lim} is minimized for $p=2.00.$ Furthermore, the entries of $\Lambda_2$ are continuous as a function of $\alpha$, which justifies the use of $\hat{\Lambda}_2 := \Lambda_2(\hat{\alpha}(\delta,2))$ as an estimator of $\Lambda_2$ when using \eqref{eq:clt}, see Appendix \ref{sec:app_lam} where we also give the specific form of $\Lambda_2.$

\begin{remark} 
We can use the above estimator of $\alpha$ to test the degree of smoothness of the paths as described above. In particular, we can test whether or not a $\mathcal{BSS}$ process is a semimartingale by testing the null hypothesis $H_0: \alpha = 0.$ In the specific case of the gamma kernel, $g(x)=x^{\alpha}e^{-\lambda},$ this test can be used to decide whether the model can be reduced to the familiar Ornstein-Uhlenbeck model. We investigate the finite sample properties of this test in Section \ref{sec:app_alpha}.
\end{remark}
\begin{remark} 
Equation \eqref{eq:clt} specifies a CLT only for $\alpha \in \left(-\frac{1}{2},\frac{1}{4}\right).$ It is possible to extend the region to include $\alpha \in \left[\frac{1}{4},\frac{1}{2}\right)$ by using gaps in sampling our observed process $\{X(i\delta)\}_{i=1}^N,$ see \cite{jose_emil_mikko_mark13}. This will, however, cause us to throw away some of the observations of $X,$ leaving us with a sparser sample and ignoring some information and we will not pursue this in the present paper.
\end{remark}

\subsubsection{Finite sample properties concerning the COF estimator}\label{sec:app_alpha}

\begin{figure*}[!t] 
\centering 
\includegraphics[scale=0.85]{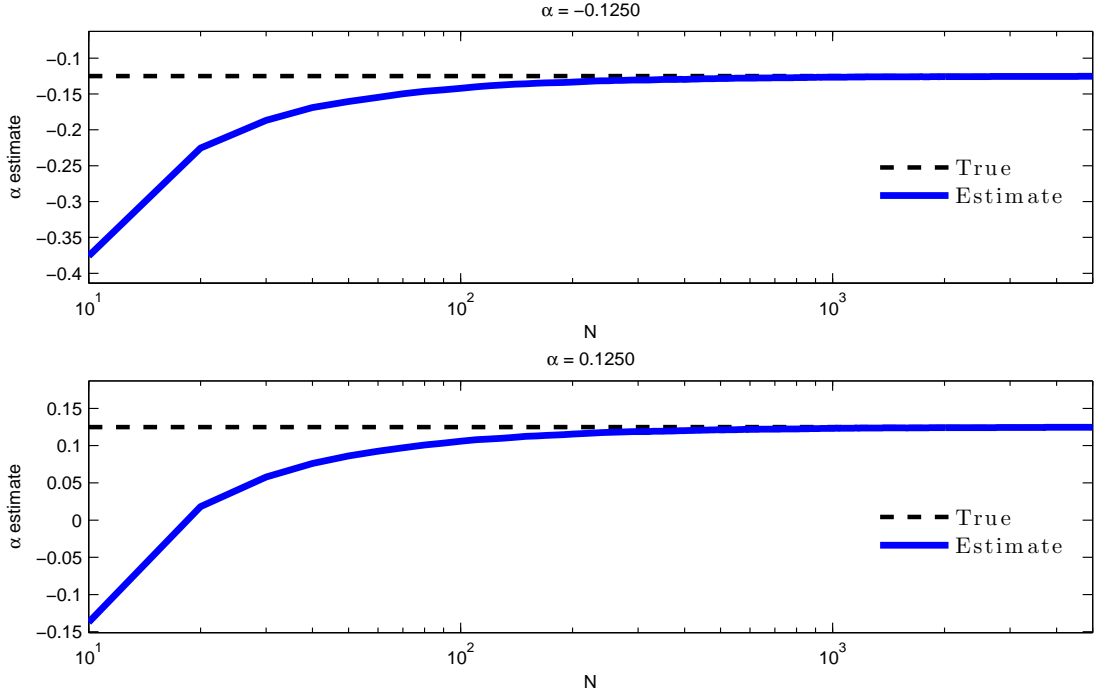}
\caption{\it Estimating $\alpha$ for an increasing number of observation, $N,$ $\alpha=-0.1250$ (top) and $\alpha=0.1250$ (bottom). Note the log-scale. The estimates have been done on the same process - that is, a process of $N_{\max} = 4000$ observations was simulated and then $\alpha$ was estimated for $N=10,20,\ldots,N_{\max}.$ Remaining parameters were  $\lambda=\sigma(t)=1$ for all $t$ and 20,000 Monte Carlo simulations was performed. We note that in both cases the estimator is biased downwards when the number of observations $N$ is small.}
\label{fig:negBias}
\end{figure*}

We now proceed to apply the discretization scheme of Section \ref{sec:simulations} to investigate the finite sample properties of the asymptotic results described above. That is, we consider how the Law of Large numbers \eqref{eq:al_lim} and Central Limit Theorem \eqref{eq:clt} of the COF estimator behave when the number of observations is finite, i.e.\ when we only observe the $\mathcal{BSS}$ process on a discrete grid of finite length. In Tables \ref{tab:rmsebias}-\ref{tab:rej3} we see investigations of how the COF estimator of $\alpha$ fares for a differing number of observations $N$ in the interval $[0,1].$ 20,000 Monte Carlo simulations have been done using the gamma kernel with $\lambda = 1$ and in three different regimes: (A) no stochastic volatility, (B) including stochastic volatility and (C) including stochastic volatility correlated with the driving Brownian motion of the $\mathcal{BSS}$ process $X$ (this phenomenon is termed \emph{leverage} in the finance literature). Table \ref{tab:rmsebias} shows that the COF estimator works satisfactorily when the number of observations are greater than  200, yielding a bias of the order $10^{-2}$ and a root mean squared error (RMSE) of around $10^{-1}.$ We note two further things. Firstly, the bias that is incurred for small values of $N$ are in all cases negative and we conclude that the estimator is \emph{biased downwards} in small samples. This is corroborated by Figure \ref{fig:negBias} where we see how the (absolute value of the) bias of the estimator decreases when increasing the number of observations of the process $X$; for small values of $N$ the estimator is severely biased downwards but as we increase the number of observations this bias vanishes. Secondly, we see that the bias and RMSE do not seem to depend on the particular regime we are in, from which we conclude that the estimator is robust to the presence of stochastic volatility and correlation effects between the stochastic volatility component and the driving noise $W.$ Next, Tables \ref{tab:rej1}-\ref{tab:rej3} investigate the CLT of the COF estimator in the three regimes by testing the null hypothesis $H_0: \alpha = \alpha_0$ for $\alpha_0 \in \{-0.1250, 0, 0.1250\}$ and various values of the true $\alpha$ used in the simulation of the $\mathcal{BSS}$ process. Again we see that for values  around $N=200$ the size of the test is satisfactory with a small upward size distortion (middle column) of the order $0.1\%.$ The power of the test (non-middle columns), however, suffers for values of $\alpha$ close to the true value unless we have many observations. We also tried varying the other parameters involved i.e.\ $\lambda$ and the stochastic volatility parameter $\beta$ but this had basically no effect on the COF estimator.  For the sake of brevity these results are not reported here but are available from the authors upon request. 

\input{TablesArxiv/Tab1.tex}

\input{TablesArxiv/Tab2.tex}

\input{TablesArxiv/Tab3.tex}

\input{TablesArxiv/Tab4.tex}

Next we investigate how the estimator performs when the number of total observations are fixed but the sampling frequency varies. Recall, that the COF estimator relies on infill asymptotics, hence a low sampling frequency could potentially be harmful to the performance. Therefore we perform simulations where the number of observations are held fixed at $N=1,000$ but the process is observed over the time period $[0,T]$ for various values of $T$ and thus different values for the step size of the observation grid, $\delta=T/N$; large $T$ corresponds to a large step size between succesive observations, i.e.\ to sampling at a low frequency. We present results using simulations with various step sizes ranging from $\delta = 1$ to $\delta = \frac{1}{200}.$ The results are shown in Table \ref{tab:lambdaTab} where we see that for large values of the parameter $\lambda,$ the estimator suffers when the step size, $\delta,$ is also large: the estimated values of $\alpha$ become biased downwards. For low values of either $\lambda$ or $\delta$ the performance is good --- in particular we conclude that the departures from the infill regime is not crucial as long as the parameter $\lambda$ is not too large.

\input{TablesArxiv/Tab5.tex}

\subsection{Estimating integrated volatility}\label{sec:sig_est}

Here we follow \cite{ole_mikko_jurgen13} and present an estimator of the integrated $p$'th power of the volatility process, $\sigma_t^{p+} := \int_0^t \sigma(s)^p ds,$ and show how we can use this to test for the presence of stochastic volatility in the process $X.$ In particular, we want to test a hypothesis of the type
\begin{align} 
H_0: \hspace{0.1cm} &\sigma(t) = \sigma_0 \hspace{0.5cm} \textnormal{for all $t\in [0,T]$} \label{eq:null} \\
H_1: \hspace{0.1cm} &\sigma(t) \neq \sigma_0 \hspace{0.5cm} \textnormal{for some $t\in [0,T]$} \nonumber
\end{align}
where $\sigma_0 \in \mathbb{R}_+.$ \\

It is well known, see e.g.\ \cite{ole_graver_jacod_mark_neil13}, that if $X=\{X(t)\}_{t \in \mathbb{R}}$ is a semimartingale, then $V_{t}^p := \sum_{i=1}^{\lfloor t/\delta \rfloor} |X(i\delta) - X((i-1)\delta)|^p \stackrel{P}{\rightarrow} m_p \sigma^{p+}$ as $\delta \rightarrow 0$ with $m_p$ as above. However, this does not hold when $X$ is not a semimartingale as is the case when $\alpha \in (-\frac{1}{2},0)\cup (0,\frac{1}{2}).$ To remedy this, we introduce the \emph{Gaussian core} of the $\mathcal{BSS}$ process, $G(t) = \int_{-\infty}^t g(t-s) dW(s)$ and define the normalization factor $c(\delta) = E[(G({\delta}) - G(0))^2]^{\frac{1}{2}}.$ Now, as shown in \cite{ole_jose_mark11} we have
\begin{align}\label{eq:ass_s}
\frac{\delta}{c(\delta)^p}V_{t}^p \stackrel{P}{\rightarrow} m_p \sigma_t^{p+}
\end{align}
as $\delta \rightarrow 0$ (see also \cite{ole-jurgen09} for the first results of this type with $p=2).$ This estimator, however, is infeasible since we in general do not know the functional form of $g$ and/or its parameters, and hence we do not know $c(\cdot).$ For this reason \cite{ole_mikko_jurgen13} introduce the \emph{realized relative power variation} (RRV) over $[0,T]$ by $\tilde{V}_{t,T}^p := \frac{V_{t}^p}{V_{T}^p}$ for $0\leq t\leq T.$ If \eqref{eq:ass_s} holds, then
\begin{align}\label{eq:RRV}
\tilde{V}_{t,T}^p \stackrel{P}{\rightarrow} \tilde{\sigma}_{t,T}^{p+} := \frac{\sigma_{t}^{p+}}{\sigma_{T}^{p+}}
\end{align}
uniformly in $t\in [0,T]$ as $\delta \rightarrow 0.$ Under some technical conditions \cite{ole_mikko_jurgen13} also provide a feasible central limit theorem for $\tilde{\sigma}_{t,T}^{p+}:$ Let
\begin{align*}
v_t(\delta) &= \frac{\lambda_{p}}{\delta \cdot m_{2p} (V_{T}^{p})^2} \left( (1-\tilde{V}_{t,T}^{p})^2 V_{t}^{2p} + (\tilde{V}_{t,T}^{p})^2(V_{T}^{2p}-V_{t}^{2p})\right).
\end{align*}
For any $t\in (0,T)$ we have
\begin{align}\label{eq:RRVclt}
\frac{\tilde{V}_{t,T}^p- \tilde{\sigma}_{t,T}^{p+}}{\sqrt{\delta v_{t}(\delta)}} \stackrel{d}{\rightarrow} N(0,1)
\end{align}
as $\delta \rightarrow 0.$  $\lambda_p = \lambda_p(\alpha)$ is a constant depending on $\alpha$ and $p$ the form of which is given in Appendix \ref{sec:app_lam}.

The RRV of equation \eqref{eq:RRV} measures the amount of accumulated volatility in $[0,t]$ compared to the total accumulated volatility in $[0,T]$ and equation \eqref{eq:RRVclt} can be used to construct confidence intervals for $\tilde{\sigma}_{t,T}^{p+}$ which allows us to test the null hypothesis that $\sigma = \{\sigma(t)\}_{t \in \mathbb{R}}$ is a constant process, i.e.\ that there is no time varying volatility present in the $\mathcal{BSS}$ process. We want to test this hypothesis, that is \eqref{eq:null} above. Under the null, we have $\tilde{\sigma}_{t,T}^{p+} = \frac{t}{T}$ and therefore
\begin{align}\label{eq:limit1}
\delta^{-\frac{1}{2}}\left(\tilde{V}_{t,T}^p - \frac{t}{T}\right) \stackrel{st}{\rightarrow} \frac{\sqrt{\lambda_p}}{m_pT} \left(W(t) - \frac{t}{T}W(T)\right),
\end{align}
where $st$ denotes \emph{stable convergence,} see \cite{renyi63} or \cite{limitTheorems} page 512 for information on this type of convergence. The right hand side of \eqref{eq:limit1} is a Brownian bridge and we utilize this to construct the hypothesis test \eqref{eq:null} by examining the distance between the empirical quantities $\tilde{V}_{1,t,T}^p$ and $\frac{t}{T}$ on the left hand side of \eqref{eq:limit1} and compare it with the critical values of the limiting distribution, i.e.\ the distribution of the distance of the properly scaled Brownian bridge in \eqref{eq:limit1}. The question becomes, which distance metric to use, as different choices will lead to different distributions of the right hand side. Two obvious choices are the $L^2$-distance and the $\sup$-distance, yielding respectively the Cramér--von Mises distribution and the Kolmogorov-Smirnov distribution. Another choice would be the $L^1$ distance, although this causes the limiting distribution to be non-standard (involving the Airy function).

\subsubsection{Finite sample properties of the test for constant volatility}\label{sec:app_SV}

\begin{figure*}[!t]
\centering
\begin{minipage}{.5\textwidth}
  \centering
 \includegraphics[scale=0.85]{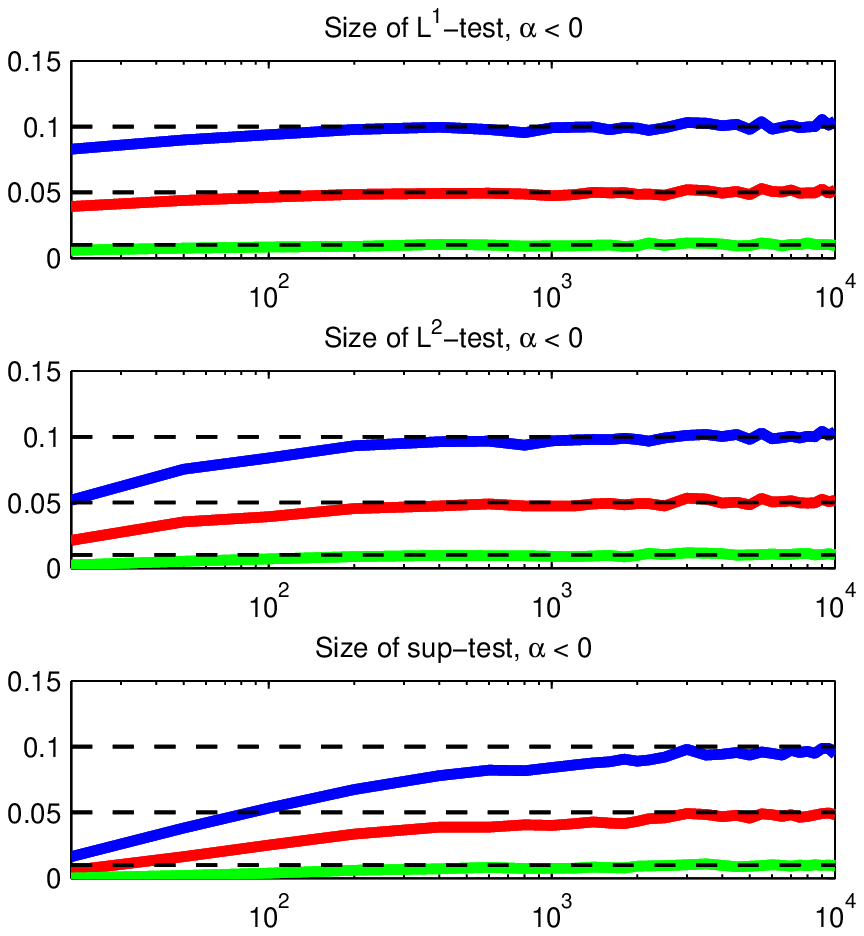}
\end{minipage}%
\begin{minipage}{.5\textwidth}
  \centering
\includegraphics[scale=0.85]{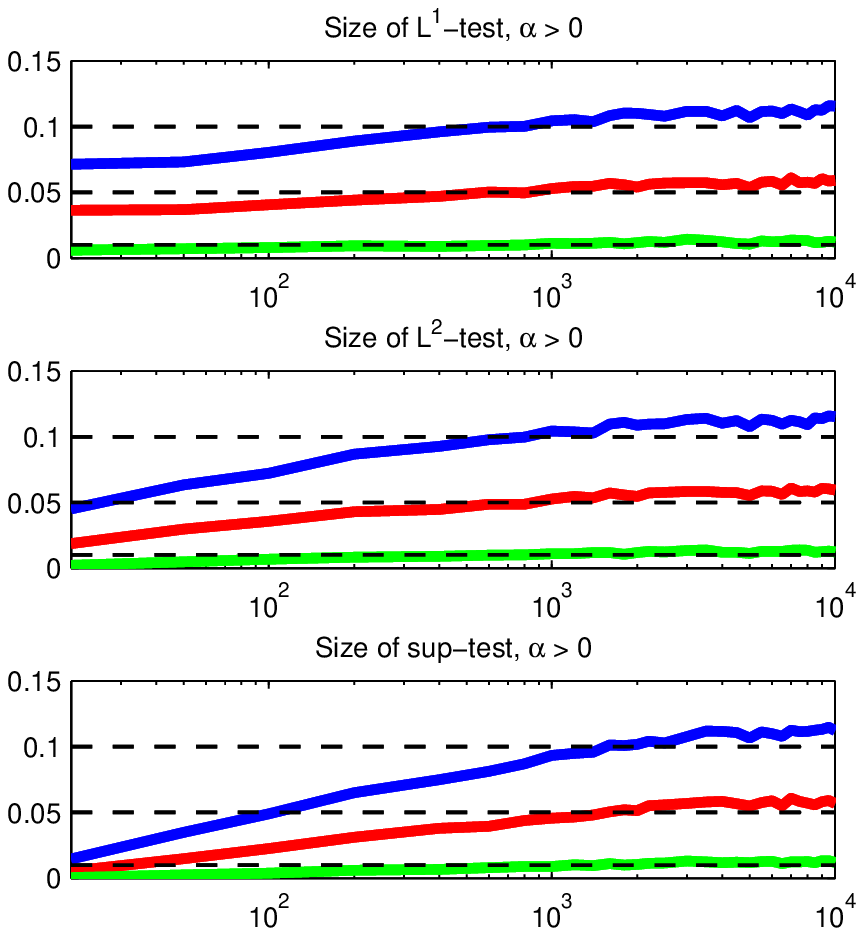}
\end{minipage}
\caption{\it Size of the test for constant volatility, $H_0 : \sigma_t = \sigma_0$ for all $t,$ with varying sample sizes, $N$ (log-scale), distance metrics and for $\alpha=-0.1250$ (left) and $\alpha=0.1250$ (right). Blue is 10\% test, red 5\% and green 1\%. $20,000$ Monte Carlo simulations.}
\label{fig:sizeRRVplot}
\end{figure*}

We explore now the finite sample properties of the test for constant volatillity, \eqref{eq:null}, in the $\mathcal{BSS}$ process by studying the convergence in \eqref{eq:limit1}. In particular, we consider the size and power of the null of constant volatility using the $L^1$-, $L^2$- and $\sup$-metrics respectively. Table \ref{tab:RRVsizeTab} shows the size of the test, i.e.\ rejection rates of the null when simulating under the null, that is when $\{\sigma_t\}_{t\in \mathbb{R}}$ is a constant process. We consider different values of the smoothness parameter $\alpha$ to investigate its effect on the test --- we similarly varied $\lambda,$ but this had practically no impact on the results, which is therefore not reported. An illustration of what is seen in the table is also given  in Figure \ref{fig:sizeRRVplot} where we plot the size of the test against the number of observations for the  three different metrics. We see, that for $\alpha =-0.1250< 0$ (left plots) the size of the $L^1$-based test is already quite accurate for around $N=100$ observations while the $L^2$ test needs around $N=200$ observations to achieve a size around the nominal value. The $\sup$-test performs markedly worse needing about $N=4000$ observations to reach the same accuracy as the two other tests. For $\alpha =0.1250 > 0$ (right plots) the relative picture between the tests is the same but the absolute picture is different: all tests need many more observations to achieve accurate sizes and even then the tests display a slight upwards size distortion. For instance it seems that the $L^1$ test needs around $N=400$ observations in this case. 

Tables \ref{tab:RRVpowPlus} and \ref{tab:a6betapow} show the power of the test, i.e.\ rejection rates of the null when simulating under the alternative when stochastic volatility is present in the $\mathcal{BSS}$ process $X.$ Simulations are presented for negative and postive values of $\alpha$ respectively and for varying values of the stochastic volatility parameter $\beta$ --- we refer to Section \ref{sec:simset} for the details on simulating the stochastic volatility process. Using the log-normality of $\sigma$ we have that $\mathbb{E}[\sigma(t)] = \exp(\tfrac{1}{4} \beta^{-1})$ and $Var(\sigma(t)) = \exp(\beta^{-1})- \exp(\tfrac{1}{2}\beta^{-1})$ so that small values of $\beta$ correspond to a large variance (and variance of variance) of $X$ and vice versa. Also, the half-life of $\{\log \sigma(t)\}_{t\in \mathbb{R}}$ is $\frac{\log (2)}{\beta}$ so that low values of $\beta$ correspond to long half-lives  (higher persistence) of the volatility of $X.$ In the tables we see what we would expect: lower $\beta$ (higher volatility of $X$) results in greater power when testing the null of constant volatility. Although the tests do show some sensitivity to the level of stochastic volatility, the power is not affected that much; Panel A ($\beta=0.05$) corresponds to a mean of $\sigma$ of $5 \cdot 10^8$ while Panel C ($\beta=5.00$) corresponds to a mean of about $1.22$ and in light of this, the differences in the size of the tests seem rather small. Additionally, comparing the two tables, we see that the power of this test is not effected by the sign of $\alpha$; the two tables --- which are for $\alpha = -0.1250$ and $\alpha = 0.1250$ respectively ---  display similar numerical values accross the board. Lastly, it is worth mentioning that the three tests all display good power in finite samples with around $80\%$ power with $N=200$ observations. The $\sup$ test has a downward bias in size for small numbers of observations and, as expected, it has higher power than the two other tests, leading us to believe that the $\sup$ test rejects more often overall as compared to the two other tests. Looking closer at the tables and Figure \ref{fig:sizeRRVplot}, we conclude from size/power considerations that the $\sup$-test is inferior to $L^1$ and $L^2$ tests, which perform almost exactly alike. Given that the $L^2$ test is based on the standard Cramér--von Mises distribution, which is widely implemented in statistical software, we recommend using the $L^2$-metric when testing the null hypothesis of constant volatility in the path of a $\mathcal{BSS}$ process.

\input{TablesArxiv/Tab6.tex}

\input{TablesArxiv/Tab7.tex}

\input{TablesArxiv/Tab8.tex}

\section{Conclusion}\label{sec:concl}

We have presented a fast and simple simulation scheme for L\'{e}vy semistationary processes and analysed the error arising from the scheme. While, a part of the error obviously stems from the truncation of the integral towards minus infinity, a more pronounced error ensues from the step function approximation when the integrating kernel has a singularity at the origin. The singularity causes the $\mathcal{LSS}$ process to be a non-semimartingale and we saw in Section \ref{sec:errorALT} how this impacts the error in the simulations. In Section \ref{sec:ass_sim} we saw an illustration of how to remedy this by sampling the process on a finer grid 
and then subsampling to get the desired path.

After providing an illustration of the simulations and the error using our main example, the gamma kernel, we applied the simulation scheme to investigate the finite sample properties of two recently developed estimators based on power variations of Brownian semistationary processes. This paper marks the first time that these estimators have been investigated in a finite sample regime and we saw that despite the infill nature of their asymptotics, the estimators performed satisfactorily when one has about $N=200$ observations per time unit $T.$ We also saw, however, that one must take caution when sampling the process infrequently; in this case, large values of the parameter $\lambda$ will cause the estimator of $\alpha$ to be downward biased.

\section*{Acknowledgements}

The research has been supported by CREATES (DNRF78), funded by the Danish National Research Foundation, 
by Aarhus University Research Foundation (project ``Stochastic and Econometric Analysis of Commodity Markets") and
by the Academy of Finland (project 258042). We would also like to thank Emil Hedevang for useful help and discussions regarding simulation of $\mathcal{LSS}$ processes.

\bibliographystyle{chicago}
\bibliography{mybib-v2}

newpage

\newpage

\section*{Appendix}

\begin{appendix}

\section{Calculating the constants $\lambda_p$ and $\Lambda_p$ of Section \ref{sec:ass}}\label{sec:app_lam}
Let $p=2,$ which is the case considered in the paper. From \cite{ole_mikko_jurgen13} we know that $\lambda_2$ is given by
\begin{align*}
\lambda(\alpha) := \lambda_2(\alpha) = 2+2\sum_{j=1}^{\infty}\rho_{\alpha}(j)^2,
\end{align*}
where for $j\geq 1,$ $\rho_{\alpha}(j)$ is the correlation function of the \emph{fractional Brownian noise} with Hurst parameter $H=\alpha +1/2,$ 
\begin{align*}
\rho_{\alpha}(j) = \frac{1}{2}\left(|j+1|^{2\alpha+1} - 2|j|^{2\alpha+1} + |j-1|^{2\alpha+1}\right).
\end{align*}
See also \cite{ole_mikko_jurgen13} for a proof of the continuity of $\alpha \mapsto \lambda_p(\alpha),$ which justifies the use of $\lambda_p(\hat{\alpha})$ as an estimator of $\lambda_p(\alpha).$ The matrix $\Lambda_p = \{\lambda_p^{ij}\}_{i,j=1}^2 $ with $\lambda_p^{ij}(\alpha) : (-\frac{1}{2},\frac{1}{4}) \rightarrow (0,\infty)$ from \eqref{eq:clt} is given by
\begin{align*}
\lambda_p^{11} &= \lim_{n\rightarrow \infty} \delta^{-1} \textnormal{Var}\left( \bar{V}_{1,1}^p(B^H)\right), \\
\lambda_p^{12} &= \lambda_p^{21} = \lim_{n\rightarrow \infty} \delta^{-1} \textnormal{Cov}\left( \bar{V}_{1,1}^p(B^H),\bar{V}_{2,1}^p (B^H)\right), \\
 \lambda_p^{22} &= \lim_{n\rightarrow \infty} \delta^{-1} \textnormal{Var}\left( \bar{V}_{2,1}^p (B^H)\right),
\end{align*}
where $\bar{V}_{k,t}^p(\cdot) = \delta c(\delta)^{-p}V_{k,t}^p(\cdot)$ and $B^H$ is fractional Brownian motion with Hurst parameter $H=\alpha + 1/2.$ For $k=1$ the above is considered with first order differences of $B^H,$ i.e.\ fractional Gaussian noise, but for $k=2$ we consider  \emph{2nd-order} differences of $B^H,$ i.e.\  $B_t^{\diamondsuit H} := B^H_t - 2 B^H_{t-1} + B^H_{t-2},$ where the correlation function of $B^{\diamondsuit H}$ is
\begin{align*}
&\rho_{\alpha}^{\diamondsuit}(j) = \frac{1}{2}(4-2^{2\alpha +1})^{-1} (-|j-2|^{2\alpha +1} +4|j-1|^{2\alpha +1} - 6j^{2\alpha +1} + 4(j+1)^{2\alpha +1}  - (j+2)^{2\alpha +1}). 
\end{align*}
As shown in \cite{jose_emil_mikko_mark13}, for $k=p=2,$ which is the case considered in this paper, the entries of $\Lambda_2$ are
\begin{align*}
\lambda_{11} &= 2 + 4\sum_{j=1}^{\infty} | \rho_{\alpha}^{\diamondsuit}(j)|^2, \\
\lambda_{12} &= 2^{2-2\alpha}(\rho_{\alpha}^{\diamondsuit}(1)+1)^2 + 2^{1-2\alpha}\sum_{j=0}^{\infty}|\rho_{\alpha}^{\diamondsuit}(j) + 2\rho_{\alpha}^{\diamondsuit}(j+1) +\rho_{\alpha}^{\diamondsuit}(j+2)|^2, \\
\lambda_{22} &= 2+2^{-4\alpha}\sum_{j=1}^{\infty}|\rho_{\alpha}^{\diamondsuit}(j-2) +4\rho_{\alpha}^{\diamondsuit}(j-1) +6\rho_{\alpha}^{\diamondsuit}(j)  +4\rho_{\alpha}^{\diamondsuit}(j+1) +\rho_{\alpha}^{\diamondsuit}(j+2)|^2.
\end{align*}
Further, following the proof of the continuity of $\alpha \mapsto \lambda_p(\alpha)$ in \cite{ole_mikko_jurgen13} it can be shown that $\alpha \mapsto \Lambda_p(\alpha)$ is continuous which justifies the use of $\Lambda_p(\hat{\alpha})$ as an estimator of $\Lambda_p(\alpha)$ in the CLT \eqref{eq:clt}.

\section{Simulation schemes}\label{sec:app_sim}

Simulating the $\mathcal{BSS},$ process $X$ without stochastic volatility is straightforward and can be done without simulation error, see \ref{sec:sim1}. Simulation of the general $\mathcal{LSS}$ process via the discretization procedure described in the paper is more involved and described in \ref{sec:sim2}.

\subsection{Exact simulation of Gaussian $\mathcal{BSS}$ process}\label{sec:sim1}

For $N$ observations in a given time period $[0,T]$ with step size $\delta = \frac{T}{N}$ do:

\begin{enumerate}
\item
	Calculate autocovariance function \eqref{eq:ACVF} for $h=0, \delta, 2\delta, \ldots, (N-1)\delta.$
\item
	Form the Toeplitz matrix  $\Sigma_{ij} := \gamma((i-j)\delta),$ $i,j=1,2,...,N$ and calculate its square root (Cholesky) matrix $F.$
\item
	Simulate a multivariate standard normal vector $Z \sim N_{N}(0,I_{N})$ and set $X = F'Z.$
\end{enumerate}

\subsection{Approximate simulation by convolution}\label{sec:sim2}

To simulate the $\mathcal{LSS}$ process with kernel function $g$ and driving Lévy process $L$ on a grid of step size $\delta$ in $[0,T]$ rewrite the integral as a sum as in Section \ref{sec:simulations} 
\begin{align*}
X(i\delta) = \int_{-\infty}^{i\delta} g(i\delta-s) \sigma(s-) dL(s) = \sum_{j=-\infty}^{i} \int_{(j-1)\delta}^{j\delta} g(i\delta-s) \sigma(s-) dL(s).
\end{align*}
Now, to approximate a path $X(i\delta)$ for $i=0,1,\ldots,\lfloor T/\Delta\rfloor$ do:

 \begin{enumerate}
    \item
      Truncate the sum towards $-\infty$ at $-M$.
    \item    
      Simulate the stochastic volatility $\sigma(j\delta)$ on a grid, $j=-M,-M+1,...,\lfloor T/\Delta\rfloor-1,$ see e.g. Section \ref{sec:simset}.
	\item
      Simulate $\Delta L_j \stackrel{d}{=} L(\delta)$ iid on a grid, $j=-M,-M+1,...,\lfloor T/\Delta\rfloor-1.$
    \item    
      Compute $G(j)=g(j\delta)$ and $\Sigma(j):=\sigma({j\delta}) \Delta L_j,$  $j=-M,-M+1,...,\lfloor T/\Delta\rfloor-1.$
    \item    
      Do discrete convolution: $Y =$ \texttt{convolution}$(G,\Sigma).$  
\item
	Select relevant values of approximate $\mathcal{BSS}$ path: $X(i\delta) \approx Y(i+M),$ $i=0,1,...,\lfloor T/\Delta\rfloor$.
\end{enumerate}

\end{appendix}

\end{document}

%% file: TablesArxiv/Tab1.tex
\begin{table*}[t]
\setlength{\tabcolsep}{0.6cm}
\caption{Estimation of $\alpha$ in three regimes}
\begin{center}
\footnotesize
\begin{tabular*}{0.99\columnwidth}{@{\extracolsep{\stretch{1}}}lr@{~~~}r@{~~~}r@{~~~}r@{~~~}r@{~~~}r@{~~~}r@{~~~}r@{~~~}r@{~~~}r@{~~~}r@{~~~}r@{~~~}r@{~~~}r@{~}} 
 \multicolumn{15}{@{}l}{\it Panel A: constant volatility} \\
\toprule
  $\alpha$ & \multicolumn{2}{c}{$-0.2500$} & & \multicolumn{2}{c}{$-0.1250$} & & \multicolumn{2}{c}{$0$} & & \multicolumn{2}{c}{$0.1250$} & & \multicolumn{2}{c}{$0.2500$} \\[0.2cm] 
 N & {\it Bias} &  {\it RMSE} & & {\it Bias} &  {\it RMSE} & & {\it Bias} &  {\it RMSE} & & {\it Bias} &  {\it RMSE} & & {\it Bias} &  {\it RMSE}  \\
\cmidrule{2-15}
$     20 $ & $ -0.088 $ & $  0.355 $& & $ -0.085 $ & $  0.338 $& & $ -0.093 $ & $  0.335 $& & $ -0.100 $ & $  0.319 $& & $ -0.101 $ & $  0.307 $\\
$     50 $ & $ -0.030 $ & $  0.213 $& & $ -0.035 $ & $  0.208 $& & $ -0.033 $ & $  0.200 $& & $ -0.037 $ & $  0.193 $& & $ -0.033 $ & $  0.182 $\\
$    100 $ & $ -0.016 $ & $  0.147 $& & $ -0.014 $ & $  0.143 $& & $ -0.018 $ & $  0.137 $& & $ -0.017 $ & $  0.134 $& & $ -0.020 $ & $  0.126 $\\
$    200 $ & $ -0.006 $ & $  0.104 $& & $ -0.009 $ & $  0.100 $& & $ -0.008 $ & $  0.098 $& & $ -0.009 $ & $  0.091 $& & $ -0.009 $ & $  0.087 $\\
$    500 $ & $ -0.005 $ & $  0.066 $& & $ -0.002 $ & $  0.063 $& & $ -0.003 $ & $  0.060 $& & $ -0.003 $ & $  0.057 $& & $ -0.005 $ & $  0.057 $\\
$   1000 $ & $ -0.001 $ & $  0.046 $& & $ -0.003 $ & $  0.045 $& & $ -0.002 $ & $  0.042 $& & $ -0.002 $ & $  0.042 $& & $ -0.002 $ & $  0.039 $\\
$   2000 $ & $ -0.001 $ & $  0.033 $& & $ -0.001 $ & $  0.032 $& & $ -0.001 $ & $  0.030 $& & $ -0.002 $ & $  0.029 $& & $ -0.001 $ & $  0.027 $\\
\bottomrule 
\end{tabular*}
\end{center}

\begin{center}
\footnotesize
\begin{tabular*}{0.99\columnwidth}{@{\extracolsep{\stretch{1}}}lr@{~~~}r@{~~~}r@{~~~}r@{~~~}r@{~~~}r@{~~~}r@{~~~}r@{~~~}r@{~~~}r@{~~~}r@{~~~}r@{~~~}r@{~~~}r@{~}} 
\multicolumn{15}{@{}l}{\it Panel B: stochastic volatility} \\
\toprule
  $\alpha$ & \multicolumn{2}{c}{$-0.2500$} & & \multicolumn{2}{c}{$-0.1250$} & & \multicolumn{2}{c}{$0$} & & \multicolumn{2}{c}{$0.1250$} & & \multicolumn{2}{c}{$0.2500$} \\[0.2cm] 
 N & {\it Bias} &  {\it RMSE} & & {\it Bias} &  {\it RMSE} & & {\it Bias} &  {\it RMSE} & & {\it Bias} &  {\it RMSE} & & {\it Bias} &  {\it RMSE}  \\
\cmidrule{2-15}
$     20 $ & $ -0.081 $ & $  0.363 $& & $ -0.094 $ & $  0.359 $& & $ -0.096 $ & $  0.348 $& & $ -0.100 $ & $  0.334 $& & $ -0.104 $ & $  0.323 $\\
$     50 $ & $ -0.027 $ & $  0.228 $& & $ -0.035 $ & $  0.222 $& & $ -0.038 $ & $  0.216 $& & $ -0.043 $ & $  0.206 $& & $ -0.041 $ & $  0.199 $\\
$    100 $ & $ -0.009 $ & $  0.162 $& & $ -0.019 $ & $  0.156 $& & $ -0.017 $ & $  0.152 $& & $ -0.018 $ & $  0.143 $& & $ -0.019 $ & $  0.140 $\\
$    200 $ & $ -0.001 $ & $  0.115 $& & $ -0.004 $ & $  0.110 $& & $ -0.008 $ & $  0.107 $& & $ -0.010 $ & $  0.104 $& & $ -0.010 $ & $  0.099 $\\
$    500 $ & $  0.003 $ & $  0.073 $& & $ -0.003 $ & $  0.070 $& & $ -0.005 $ & $  0.069 $& & $ -0.004 $ & $  0.064 $& & $ -0.004 $ & $  0.062 $\\
$   1000 $ & $  0.004 $ & $  0.052 $& & $  0.001 $ & $  0.050 $& & $ -0.001 $ & $  0.048 $& & $ -0.003 $ & $  0.047 $& & $ -0.002 $ & $  0.044 $\\
$   2000 $ & $  0.005 $ & $  0.038 $& & $  0.001 $ & $  0.035 $& & $ -0.002 $ & $  0.035 $& & $ -0.001 $ & $  0.034 $& & $ -0.001 $ & $  0.031 $\\
\bottomrule 
\end{tabular*}
\end{center}

\begin{center}
\footnotesize
\begin{tabular*}{0.99\columnwidth}{@{\extracolsep{\stretch{1}}}lr@{~~~}r@{~~~}r@{~~~}r@{~~~}r@{~~~}r@{~~~}r@{~~~}r@{~~~}r@{~~~}r@{~~~}r@{~~~}r@{~~~}r@{~~~}r@{~}} 
 \multicolumn{15}{@{}l}{\it Panel C: stochastic volatility correlated with the driving noise} \\
\toprule
  $\alpha$ & \multicolumn{2}{c}{$-0.2500$} & & \multicolumn{2}{c}{$-0.1250$} & & \multicolumn{2}{c}{$0$} & & \multicolumn{2}{c}{$0.1250$} & & \multicolumn{2}{c}{$0.2500$} \\[0.2cm] 
 N & {\it Bias} &  {\it RMSE} & & {\it Bias} &  {\it RMSE} & & {\it Bias} &  {\it RMSE} & & {\it Bias} &  {\it RMSE} & & {\it Bias} &  {\it RMSE}  \\
\cmidrule{2-15}  
$     20 $ & $ -0.093 $ & $  0.367 $& & $ -0.100 $ & $  0.359 $& & $ -0.104 $ & $  0.351 $& & $ -0.107 $ & $  0.342 $& & $ -0.112 $ & $  0.332 $\\
$     50 $ & $ -0.026 $ & $  0.228 $& & $ -0.032 $ & $  0.222 $& & $ -0.035 $ & $  0.216 $& & $ -0.038 $ & $  0.208 $& & $ -0.040 $ & $  0.199 $\\
$    100 $ & $ -0.012 $ & $  0.162 $& & $ -0.018 $ & $  0.157 $& & $ -0.019 $ & $  0.152 $& & $ -0.019 $ & $  0.146 $& & $ -0.020 $ & $  0.139 $\\
$    200 $ & $ -0.003 $ & $  0.114 $& & $ -0.008 $ & $  0.110 $& & $ -0.010 $ & $  0.106 $& & $ -0.010 $ & $  0.101 $& & $ -0.010 $ & $  0.096 $\\
$    500 $ & $  0.001 $ & $  0.073 $& & $ -0.004 $ & $  0.070 $& & $ -0.004 $ & $  0.068 $& & $ -0.004 $ & $  0.065 $& & $ -0.004 $ & $  0.062 $\\
$   1000 $ & $  0.004 $ & $  0.051 $& & $ -0.001 $ & $  0.049 $& & $ -0.002 $ & $  0.047 $& & $ -0.002 $ & $  0.046 $& & $ -0.003 $ & $  0.044 $\\
$   2000 $ & $  0.005 $ & $  0.037 $& & $  0.000 $ & $  0.035 $& & $ -0.001 $ & $  0.034 $& & $ -0.001 $ & $  0.032 $& & $ -0.001 $ & $  0.031 $\\
\bottomrule 
\end{tabular*}
\end{center}
{\footnotesize \it Bias and root mean squared error of the estimator of $\alpha$ for varying values of true $\alpha$ and in the three regimes; no stochastic volatility (Panel A), including stochastic volatility (Panel B) and including stochastic volatility which is also correlated with the driving noise of the $\mathcal{BSS}$ process (Panel C). $\lambda=1, \beta=5, \rho=-0.5.$ $20,000$ Monte Carlo simulations.}
\label{tab:rmsebias}
\end{table*}

%% file: TablesArxiv/Tab2.tex
\begin{table*}[t]
\caption{Rejection rates of $H_0: \alpha =\alpha_0$ with constant volatility}
\setlength{\tabcolsep}{0.6cm}
\begin{center}
\footnotesize
\begin{tabular*}{0.99\columnwidth}{@{\extracolsep{\stretch{1}}}lr@{~~~}r@{~~~}r@{~~~}r@{~~~}r@{~~~}r@{~~~}r@{~~~}r@{~~~}r@{~~~}r@{~~~}r@{~~~}r@{~~~}r@{~~~}r@{~}}
\multicolumn{12}{@{}l}{\it  Panel A: $\alpha_0 = -0.1250$} \\
\toprule
& \multicolumn{11}{c}{True value of $\alpha$} \\[0.2cm]
 N &  $ -0.4950$ & $-0.3750$ &  $-0.2500$ & $-0.1875$ & & $-0.1250$ & & $-0.0625$ &  $0$   & $0.1250$ &   $0.2500$   \\
\cmidrule(r){2-12}
     20  &  0.333 &  0.252 &  0.154 &  0.125 &  &  0.103 &  &  0.088 &  0.075 &  0.085 &  0.143\\
     50  &  0.518 &  0.311 &  0.159 &  0.099 &  &  0.074 &  &  0.063 &  0.082 &  0.189 &  0.408\\
    100  &  0.755 &  0.468 &  0.199 &  0.103 &  &  0.059 &  &  0.063 &  0.119 &  0.384 &  0.738\\
    200  &  0.954 &  0.720 &  0.263 &  0.107 &  &  0.053 &  &  0.090 &  0.216 &  0.707 &  0.970\\
    500  &  1.000 &  0.967 &  0.530 &  0.198 &  &  0.057 &  &  0.156 &  0.496 &  0.979 &  1.000\\
   1000  &  1.000 &  0.999 &  0.799 &  0.300 &  &  0.047 &  &  0.278 &  0.806 &  1.000 &  1.000\\
   2000  &  1.000 &  1.000 &  0.973 &  0.530 &  &  0.055 &  &  0.498 &  0.980 &  1.000 &  1.000\\
\bottomrule
 \end{tabular*}
\end{center}

\begin{center}
\footnotesize
\begin{tabular*}{0.99\columnwidth}{@{\extracolsep{\stretch{1}}}lr@{~~~}r@{~~~}r@{~~~}r@{~~~}r@{~~~}r@{~~~}r@{~~~}r@{~~~}r@{~~~}r@{~~~}r@{~~~}r@{~~~}r@{~~~}r@{~}}
\multicolumn{12}{@{}l}{\it Panel B: $\alpha_0 = 0$} \\
\toprule
& \multicolumn{11}{c}{True value of $\alpha$} \\[0.2cm]
 N & $-0.3750$ & $-0.2500$ & $ -0.1250 $& $ -0.0625 $& & $0$ &  & $0.0625$ & $0.1250$ &   $0.2500$ & $0.3750$   \\
\cmidrule(r){2-12}
     20  &  0.373 &  0.250 &  0.165 &  0.129 &  &  0.101 &  &  0.076 &  0.068 &  0.084 &  0.139\\
     50  &  0.556 &  0.339 &  0.175 &  0.099 &  &  0.067 &  &  0.061 &  0.073 &  0.200 &  0.450\\
    100  &  0.798 &  0.508 &  0.201 &  0.099 &  &  0.057 &  &  0.064 &  0.120 &  0.409 &  0.795\\
    200  &  0.969 &  0.751 &  0.283 &  0.129 &  &  0.055 &  &  0.076 &  0.231 &  0.730 &  0.981\\
    500  &  1.000 &  0.980 &  0.559 &  0.197 &  &  0.062 &  &  0.160 &  0.527 &  0.990 &  1.000\\
   1000  &  1.000 &  1.000 &  0.828 &  0.323 &  &  0.049 &  &  0.275 &  0.838 &  1.000 &  1.000\\
   2000  &  1.000 &  1.000 &  0.981 &  0.555 &  &  0.049 &  &  0.533 &  0.986 &  1.000 &  1.000\\
\bottomrule
 \end{tabular*}
\end{center}

\begin{center}
\footnotesize
\begin{tabular*}{0.99\columnwidth}{@{\extracolsep{\stretch{1}}}lr@{~~~}r@{~~~}r@{~~~}r@{~~~}r@{~~~}r@{~~~}r@{~~~}r@{~~~}r@{~~~}r@{~~~}r@{~~~}r@{~~~}r@{~~~}r@{~}}
\multicolumn{12}{@{}l}{\it Panel C: $\alpha_0 = 0.1250$}  \\
\toprule
& \multicolumn{11}{c}{True value of $\alpha$} \\[0.2cm]
 N  &  $-0.2500$ &  $-0.1250$ & $0$  & $0.0625$ & &$0.1250$ & & $0.1875$ &   $0.2500$ & $0.3750$  &$0.4950$   \\
\cmidrule(r){2-12}
     20  &  0.385 &  0.268 &  0.171 &  0.132 &  &  0.094 &  &  0.086 &  0.068 &  0.077 &  0.146\\
     50  &  0.605 &  0.371 &  0.169 &  0.113 &  &  0.064 &  &  0.067 &  0.078 &  0.225 &  0.455\\
    100  &  0.825 &  0.534 &  0.201 &  0.098 &  &  0.063 &  &  0.069 &  0.123 &  0.447 &  0.819\\
    200  &  0.979 &  0.777 &  0.334 &  0.131 &  &  0.056 &  &  0.086 &  0.249 &  0.775 &  0.987\\
    500  &  1.000 &  0.988 &  0.594 &  0.217 &  &  0.054 &  &  0.191 &  0.561 &  0.993 &  1.000\\
   1000  &  1.000 &  1.000 &  0.864 &  0.358 &  &  0.054 &  &  0.312 &  0.872 &  1.000 &  1.000\\
   2000  &  1.000 &  1.000 &  0.990 &  0.588 &  &  0.052 &  &  0.589 &  0.994 &  1.000 &  1.000\\
\bottomrule
 \end{tabular*}
\end{center}
{\footnotesize  \it Simulations of the test $H_0: \alpha = \alpha_0$ at a nominal level of $5\%$ for $\alpha_0= -0.1250$ (Panel A), $\alpha_0 = 0$ (Panel B), $\alpha_0= 0.1250$ (Panel C) and varying values of the true $\alpha.$ Numbers are rejection rates of the null when simulating the underlying process using the value of $\alpha$ given in the top row; thus mid columns correspond to the size of the test while non-mid columns correspond to power. Simulations are with constant volatility, $\sigma_t=1$ for all $t$ and with $\lambda=1.$ $20,000$ Monte Carlo simulations.}
\label{tab:rej1}
\end{table*}

%% file: TablesArxiv/Tab3.tex
\begin{table*}[t]
\caption{Rejection rates of $H_0: \alpha =\alpha_0$ with stochastic volatility}
\setlength{\tabcolsep}{0.6cm}
\begin{center}
\footnotesize
\begin{tabular*}{0.99\columnwidth}{@{\extracolsep{\stretch{1}}}lr@{~~~}r@{~~~}r@{~~~}r@{~~~}r@{~~~}r@{~~~}r@{~~~}r@{~~~}r@{~~~}r@{~~~}r@{~~~}r@{~~~}r@{~~~}r@{~}}
\multicolumn{12}{@{}l}{\it Panel A: $\alpha_0 = -0.1250$} \\
\toprule
& \multicolumn{11}{c}{True value of $\alpha$} \\[0.2cm]
 N &  $ -0.4950$ & $-0.3750$ &  $-0.2500$ & $-0.1875$ & & $-0.1250$ & & $-0.0625$ &  $0$   & $0.1250$ &   $0.2500$   \\
\cmidrule(r){2-12}
     20  &  0.251 &  0.201 &  0.139 &  0.114 &  &  0.095 &  &  0.080 &  0.069 &  0.095 &  0.142\\
     50  &  0.334 &  0.243 &  0.140 &  0.098 &  &  0.073 &  &  0.069 &  0.077 &  0.169 &  0.363\\
    100  &  0.493 &  0.329 &  0.146 &  0.086 &  &  0.057 &  &  0.072 &  0.119 &  0.335 &  0.676\\
    200  &  0.734 &  0.514 &  0.213 &  0.100 &  &  0.056 &  &  0.086 &  0.194 &  0.602 &  0.917\\
    500  &  0.973 &  0.859 &  0.382 &  0.140 &  &  0.047 &  &  0.135 &  0.402 &  0.937 &  1.000\\
   1000  &  1.000 &  0.988 &  0.624 &  0.196 &  &  0.049 &  &  0.244 &  0.717 &  0.999 &  1.000\\
   2000  &  1.000 &  1.000 &  0.891 &  0.380 &  &  0.049 &  &  0.445 &  0.940 &  1.000 &  1.000\\
\bottomrule
 \end{tabular*}
\end{center}

\begin{center}
\footnotesize
\begin{tabular*}{0.99\columnwidth}{@{\extracolsep{\stretch{1}}}lr@{~~~}r@{~~~}r@{~~~}r@{~~~}r@{~~~}r@{~~~}r@{~~~}r@{~~~}r@{~~~}r@{~~~}r@{~~~}r@{~~~}r@{~~~}r@{~}}
\multicolumn{12}{@{}l}{\it Panel B: $\alpha_0 = 0$} \\
\toprule
& \multicolumn{11}{c}{True value of $\alpha$} \\[0.2cm]
 N & $-0.3750$ & $-0.2500$ & $ -0.1250 $& $ -0.0625 $& & $0$ &  & $0.0625$ & $0.1250$ &   $0.2500$ & $0.3750$   \\
\cmidrule(r){2-12}
     20  &  0.302 &  0.215 &  0.150 &  0.121 &  &  0.093 &  &  0.074 &  0.064 &  0.087 &  0.143\\
     50  &  0.450 &  0.290 &  0.159 &  0.103 &  &  0.079 &  &  0.069 &  0.082 &  0.174 &  0.391\\
    100  &  0.647 &  0.411 &  0.172 &  0.090 &  &  0.055 &  &  0.066 &  0.113 &  0.367 &  0.712\\
    200  &  0.868 &  0.625 &  0.248 &  0.113 &  &  0.060 &  &  0.083 &  0.201 &  0.630 &  0.941\\
    500  &  0.999 &  0.934 &  0.468 &  0.175 &  &  0.049 &  &  0.136 &  0.436 &  0.948 &  1.000\\
   1000  &  1.000 &  0.997 &  0.718 &  0.240 &  &  0.047 &  &  0.251 &  0.752 &  0.999 &  1.000\\
   2000  &  1.000 &  1.000 &  0.948 &  0.451 &  &  0.052 &  &  0.446 &  0.955 &  1.000 &  1.000\\
\bottomrule
 \end{tabular*}
\end{center}

\begin{center}
\footnotesize
\begin{tabular*}{0.99\columnwidth}{@{\extracolsep{\stretch{1}}}lr@{~~~}r@{~~~}r@{~~~}r@{~~~}r@{~~~}r@{~~~}r@{~~~}r@{~~~}r@{~~~}r@{~~~}r@{~~~}r@{~~~}r@{~~~}r@{~}}
\multicolumn{12}{@{}l}{\it Panel C: $\alpha_0 = 0.1250$}  \\
\toprule
& \multicolumn{11}{c}{True value of $\alpha$} \\[0.2cm]
 N  &  $-0.2500$ &  $-0.1250$ & $0$  & $0.0625$ & &$0.1250$ & & $0.1875$ &   $0.2500$ & $0.3750$  &$0.4950$   \\
\cmidrule(r){2-12} 
     20  &  0.452 &  0.361 &  0.246 &  0.164 &  &  0.099 &  &  0.080 &  0.069 &  0.084 &  0.138\\
     50  &  0.696 &  0.514 &  0.333 &  0.171 &  &  0.077 &  &  0.069 &  0.081 &  0.185 &  0.394\\
    100  &  0.882 &  0.724 &  0.469 &  0.186 &  &  0.059 &  &  0.067 &  0.121 &  0.393 &  0.739\\
    200  &  0.992 &  0.935 &  0.686 &  0.272 &  &  0.057 &  &  0.088 &  0.207 &  0.667 &  0.955\\
    500  &  1.000 &  1.000 &  0.969 &  0.521 &  &  0.051 &  &  0.141 &  0.460 &  0.965 &  1.000\\
   1000  &  1.000 &  0.998 &  0.758 &  0.271 &  &  0.046 &  &  0.273 &  0.788 &  1.000 &  1.000\\
   2000  &  1.000 &  1.000 &  1.000 &  0.964 &  &  0.057 &  &  0.469 &  0.968 &  1.000 &  1.000\\
\bottomrule
 \end{tabular*}
\end{center}
{\footnotesize \it Simulations of the test $H_0: \alpha = \alpha_0$ at a nominal level of $5\%$ for $\alpha_0= -0.1250$ (Panel A), $\alpha_0 = 0$ (Panel B), $\alpha_0= 0.1250$ (Panel C) and varying values of the true $\alpha.$ Numbers are rejection rates of the null when simulating the underlying process using the value of $\alpha$ given in the top row; thus mid columns correspond to the size of the test while non-mid columns correspond to power. Simulations are with stochastic volatility with parameter  $\beta=5$ and $\lambda=1.$ $20,000$ Monte Carlo simulations.}
\label{tab:rej2}
\end{table*}

%% file: TablesArxiv/Tab4.tex
\begin{table*}[t]
\setlength{\tabcolsep}{0.6cm}
\caption{Rejection rates of $H_0: \alpha =\alpha_0$ with stochastic volatility correlated with the driving noise}
\begin{center}
\footnotesize
\begin{tabular*}{0.99\columnwidth}{@{\extracolsep{\stretch{1}}}lr@{~~~}r@{~~~}r@{~~~}r@{~~~}r@{~~~}r@{~~~}r@{~~~}r@{~~~}r@{~~~}r@{~~~}r@{~~~}r@{~~~}r@{~~~}r@{~}}
\multicolumn{12}{@{}l}{\it Panel A: $\alpha_0 = -0.1250$} \\
\toprule
& \multicolumn{11}{c}{True value of $\alpha$} \\[0.2cm]
 N &  $ -0.4950$ & $-0.3750$ &  $-0.2500$ & $-0.1875$ & & $-0.1250$ & & $-0.0625$ &  $0$   & $0.1250$ &   $0.2500$   \\
\cmidrule(r){2-12}
     20  &  0.256 &  0.202 &  0.138 &  0.114 &  &  0.097 &  &  0.075 &  0.070 &  0.089 &  0.144\\
     50  &  0.342 &  0.246 &  0.143 &  0.102 &  &  0.075 &  &  0.070 &  0.082 &  0.172 &  0.367\\
    100  &  0.484 &  0.330 &  0.153 &  0.095 &  &  0.060 &  &  0.065 &  0.112 &  0.337 &  0.673\\
    200  &  0.736 &  0.525 &  0.209 &  0.104 &  &  0.059 &  &  0.085 &  0.195 &  0.600 &  0.924\\
    500  &  0.981 &  0.863 &  0.386 &  0.142 &  &  0.052 &  &  0.142 &  0.425 &  0.945 &  1.000\\
   1000  &  0.999 &  0.988 &  0.628 &  0.203 &  &  0.051 &  &  0.255 &  0.719 &  1.000 &  1.000\\
   2000  &  1.000 &  1.000 &  0.894 &  0.376 &  &  0.056 &  &  0.453 &  0.945 &  1.000 &  1.000\\
\bottomrule
 \end{tabular*}
\end{center}

\begin{center}
\footnotesize
\begin{tabular*}{0.99\columnwidth}{@{\extracolsep{\stretch{1}}}lr@{~~~}r@{~~~}r@{~~~}r@{~~~}r@{~~~}r@{~~~}r@{~~~}r@{~~~}r@{~~~}r@{~~~}r@{~~~}r@{~~~}r@{~~~}r@{~}}
\multicolumn{12}{@{}l}{\it Panel B: $\alpha_0 = 0$} \\
\toprule
& \multicolumn{11}{c}{True value of $\alpha$} \\[0.2cm]
 N & $-0.3750$ & $-0.2500$ & $ -0.1250 $& $ -0.0625 $& & $0$ &  & $0.0625$ & $0.1250$ &   $0.2500$ & $0.3750$   \\
\cmidrule(r){2-12}
     20  &  0.296 &  0.223 &  0.148 &  0.117 &  &  0.093 &  &  0.076 &  0.070 &  0.093 &  0.139\\
     50  &  0.450 &  0.298 &  0.155 &  0.106 &  &  0.077 &  &  0.067 &  0.082 &  0.174 &  0.385\\
    100  &  0.642 &  0.409 &  0.181 &  0.101 &  &  0.055 &  &  0.063 &  0.117 &  0.366 &  0.718\\
    200  &  0.872 &  0.629 &  0.242 &  0.114 &  &  0.063 &  &  0.084 &  0.203 &  0.632 &  0.947\\
    500  &  0.999 &  0.940 &  0.473 &  0.171 &  &  0.050 &  &  0.145 &  0.438 &  0.955 &  1.000\\
   1000  &  1.000 &  0.995 &  0.722 &  0.254 &  &  0.047 &  &  0.275 &  0.747 &  1.000 &  1.000\\
   2000  &  1.000 &  1.000 &  0.948 &  0.448 &  &  0.051 &  &  0.455 &  0.956 &  1.000 &  1.000\\
\bottomrule
 \end{tabular*}
\end{center}

\begin{center}
\footnotesize
\begin{tabular*}{0.99\columnwidth}{@{\extracolsep{\stretch{1}}}lr@{~~~}r@{~~~}r@{~~~}r@{~~~}r@{~~~}r@{~~~}r@{~~~}r@{~~~}r@{~~~}r@{~~~}r@{~~~}r@{~~~}r@{~~~}r@{~}}
\multicolumn{12}{@{}l}{\it Panel C: $\alpha_0 = 0.1250$}  \\
\toprule
& \multicolumn{11}{c}{True value of $\alpha$} \\[0.2cm]
 N  &  $-0.2500$ &  $-0.1250$ & $0$  & $0.0625$ & &$0.1250$ & & $0.1875$ &   $0.2500$ & $0.3750$  &$0.4950$   \\
\cmidrule(r){2-12}
     20  &  0.354 &  0.249 &  0.164 &  0.126 &  &  0.096 &  &  0.080 &  0.071 &  0.084 &  0.134\\
     50  &  0.523 &  0.335 &  0.169 &  0.117 &  &  0.076 &  &  0.067 &  0.081 &  0.179 &  0.404\\
    100  &  0.728 &  0.461 &  0.191 &  0.102 &  &  0.057 &  &  0.070 &  0.122 &  0.393 &  0.741\\
    200  &  0.935 &  0.692 &  0.271 &  0.119 &  &  0.058 &  &  0.088 &  0.216 &  0.662 &  0.965\\
    500  &  1.000 &  0.970 &  0.516 &  0.191 &  &  0.053 &  &  0.148 &  0.469 &  0.971 &  1.000\\
   1000  &  1.000 &  0.998 &  0.768 &  0.281 &  &  0.046 &  &  0.288 &  0.784 &  1.000 &  1.000\\
   2000  &  1.000 &  1.000 &  0.964 &  0.500 &  &  0.056 &  &  0.479 &  0.971 &  1.000 &  1.000\\
\bottomrule
 \end{tabular*}
\end{center}
{\footnotesize \it  Simulations of the test $H_0: \alpha = \alpha_0$ at a nominal level of $5\%$ for $\alpha_0= -0.1250$ (Panel A), $\alpha_0 = 0$ (Panel B), $\alpha_0= 0.1250$ (Panel C) and varying values of the true $\alpha.$ Numbers are rejection rates of the null when simulating the underlying process using the value of $\alpha$ given in the top row; thus mid columns correspond to the size of the test while non-mid columns correspond to power. Simulations are including stochastic volatility correlated with the driving process of the $\mathcal{BSS}$ process. The stochastic volatility parameter is $\beta=5$ and the correlation coefficient $\rho = -0.5.$ $\lambda=1.$ $20,000$ Monte Carlo simulations.}
\label{tab:rej3}
\end{table*}

%% file: TablesArxiv/Tab5.tex
\begin{table*}[t]
\caption{Estimation of $\alpha$ when sampling infrequently}
\setlength{\tabcolsep}{0.6cm}
\begin{center}
\footnotesize
\begin{tabular*}{0.99\columnwidth}{@{\extracolsep{\stretch{1}}}lr@{~~~}r@{~~~}r@{~~~}r@{~~~}r@{~~~}r@{~~~}r@{~~~}r@{~~~}r@{~~~}r@{~~~}r@{~~~}r@{~~~}r@{~~~}r@{~}} 
\multicolumn{15}{@{}l}{\it Panel A: $\lambda = 0.01$} \\
\toprule
 $\alpha$ & \multicolumn{2}{c}{$-0.2500$} & & \multicolumn{2}{c}{$-0.1250$} & & \multicolumn{2}{c}{$0$} & & \multicolumn{2}{c}{$0.1250$} & & \multicolumn{2}{c}{$0.2500$} \\[0.2cm]
$\delta$ & {\it Bias} &  {\it RMSE} & & {\it Bias} &  {\it RMSE} & & {\it Bias} &  {\it RMSE} & & {\it Bias} &  {\it RMSE} & & {\it Bias} &  {\it RMSE}  \\
\cmidrule{2-15}
   1.00 & $ -0.001 $ & $  0.046 $ &  & $ -0.001 $ & $  0.045 $ &  & $ -0.001 $ & $  0.043 $ &  & $ -0.002 $ & $  0.041 $ &  & $ -0.002 $ & $  0.039 $\\
   0.50 & $ -0.002 $ & $  0.047 $ &  & $ -0.001 $ & $  0.044 $ &  & $ -0.001 $ & $  0.043 $ &  & $ -0.002 $ & $  0.041 $ &  & $ -0.002 $ & $  0.039 $\\
   0.20 & $ -0.001 $ & $  0.046 $ &  & $ -0.002 $ & $  0.044 $ &  & $ -0.002 $ & $  0.043 $ &  & $ -0.002 $ & $  0.041 $ &  & $ -0.002 $ & $  0.039 $\\
   0.10 & $ -0.001 $ & $  0.046 $ &  & $ -0.001 $ & $  0.044 $ &  & $ -0.002 $ & $  0.043 $ &  & $ -0.001 $ & $  0.041 $ &  & $ -0.002 $ & $  0.039 $\\
   0.05 & $ -0.002 $ & $  0.046 $ &  & $ -0.002 $ & $  0.044 $ &  & $ -0.001 $ & $  0.043 $ &  & $ -0.002 $ & $  0.041 $ &  & $ -0.002 $ & $  0.039 $\\
   0.02 & $ -0.002 $ & $  0.046 $ &  & $ -0.002 $ & $  0.044 $ &  & $ -0.002 $ & $  0.043 $ &  & $ -0.002 $ & $  0.041 $ &  & $ -0.002 $ & $  0.039 $\\
\bottomrule 
\end{tabular*}
\end{center}
\begin{center}
\footnotesize
\begin{tabular*}{0.99\columnwidth}{@{\extracolsep{\stretch{1}}}lr@{~~~}r@{~~~}r@{~~~}r@{~~~}r@{~~~}r@{~~~}r@{~~~}r@{~~~}r@{~~~}r@{~~~}r@{~~~}r@{~~~}r@{~~~}r@{~}} 
\multicolumn{15}{@{}l}{\it Panel B: $\lambda = 0.1$}\\
\toprule
 $\alpha$ & \multicolumn{2}{c}{$-0.2500$} & & \multicolumn{2}{c}{$-0.1250$} & & \multicolumn{2}{c}{$0$} & & \multicolumn{2}{c}{$0.1250$} & & \multicolumn{2}{c}{$0.2500$} \\[0.2cm]
$\delta$ & {\it Bias} &  {\it RMSE} & & {\it Bias} &  {\it RMSE} & & {\it Bias} &  {\it RMSE} & & {\it Bias} &  {\it RMSE} & & {\it Bias} &  {\it RMSE}  \\
\cmidrule{2-15}
   1.00 & $ -0.004 $ & $  0.046 $ &  & $ -0.006 $ & $  0.045 $ &  & $ -0.008 $ & $  0.044 $ &  & $ -0.010 $ & $  0.042 $ &  & $ -0.014 $ & $  0.042 $\\
   0.50 & $ -0.002 $ & $  0.046 $ &  & $ -0.003 $ & $  0.044 $ &  & $ -0.004 $ & $  0.043 $ &  & $ -0.004 $ & $  0.041 $ &  & $ -0.005 $ & $  0.039 $\\
   0.20 & $ -0.002 $ & $  0.046 $ &  & $ -0.002 $ & $  0.044 $ &  & $ -0.002 $ & $  0.043 $ &  & $ -0.002 $ & $  0.040 $ &  & $ -0.002 $ & $  0.039 $\\
   0.10 & $ -0.002 $ & $  0.046 $ &  & $ -0.002 $ & $  0.045 $ &  & $ -0.001 $ & $  0.043 $ &  & $ -0.002 $ & $  0.041 $ &  & $ -0.002 $ & $  0.039 $\\
   0.05 & $ -0.002 $ & $  0.046 $ &  & $ -0.002 $ & $  0.045 $ &  & $ -0.002 $ & $  0.043 $ &  & $ -0.002 $ & $  0.041 $ &  & $ -0.002 $ & $  0.039 $\\
   0.02 & $ -0.002 $ & $  0.046 $ &  & $ -0.002 $ & $  0.044 $ &  & $ -0.001 $ & $  0.042 $ &  & $ -0.002 $ & $  0.041 $ &  & $ -0.002 $ & $  0.039 $\\
\bottomrule 
\end{tabular*}
\end{center}
\begin{center}
\footnotesize
\begin{tabular*}{0.99\columnwidth}{@{\extracolsep{\stretch{1}}}lr@{~~~}r@{~~~}r@{~~~}r@{~~~}r@{~~~}r@{~~~}r@{~~~}r@{~~~}r@{~~~}r@{~~~}r@{~~~}r@{~~~}r@{~~~}r@{~}} 
\multicolumn{15}{@{}l}{\it Panel C: $\lambda = 1$}\\
\toprule
 $\alpha$ & \multicolumn{2}{c}{$-0.2500$} & & \multicolumn{2}{c}{$-0.1250$} & & \multicolumn{2}{c}{$0$} & & \multicolumn{2}{c}{$0.1250$} & & \multicolumn{2}{c}{$0.2500$} \\[0.2cm]
$\delta$ & {\it Bias} &  {\it RMSE} & & {\it Bias} &  {\it RMSE} & & {\it Bias} &  {\it RMSE} & & {\it Bias} &  {\it RMSE} & & {\it Bias} &  {\it RMSE}  \\
\cmidrule{2-15}
   1.00 & $ -0.111 $ & $  0.121 $ &  & $ -0.163 $ & $  0.169 $ &  & $ -0.214 $ & $  0.219 $ &  & $ -0.266 $ & $  0.270 $ &  & $ -0.318 $ & $  0.321 $\\
   0.50 & $ -0.045 $ & $  0.065 $ &  & $ -0.067 $ & $  0.081 $ &  & $ -0.091 $ & $  0.101 $ &  & $ -0.116 $ & $  0.123 $ &  & $ -0.143 $ & $  0.149 $\\
   0.20 & $ -0.011 $ & $  0.048 $ &  & $ -0.016 $ & $  0.048 $ &  & $ -0.023 $ & $  0.049 $ &  & $ -0.030 $ & $  0.051 $ &  & $ -0.039 $ & $  0.056 $\\
   0.10 & $ -0.004 $ & $  0.046 $ &  & $ -0.006 $ & $  0.045 $ &  & $ -0.008 $ & $  0.044 $ &  & $ -0.010 $ & $  0.042 $ &  & $ -0.014 $ & $  0.041 $\\
   0.05 & $ -0.002 $ & $  0.046 $ &  & $ -0.003 $ & $  0.045 $ &  & $ -0.003 $ & $  0.043 $ &  & $ -0.004 $ & $  0.041 $ &  & $ -0.005 $ & $  0.039 $\\
   0.02 & $ -0.001 $ & $  0.046 $ &  & $ -0.002 $ & $  0.045 $ &  & $ -0.002 $ & $  0.043 $ &  & $ -0.002 $ & $  0.041 $ &  & $ -0.002 $ & $  0.039 $\\
\bottomrule 
\end{tabular*}
\end{center}
{\footnotesize \it Investigation of the bias incurred when the $\mathcal{BSS}$ process is sampled infrequently. The number of observations is held fixed at $N=1000$ but the step size between successive observations $\delta$ varies. 20,000 Monte Carlo simulations.} 
\label{tab:lambdaTab}
\end{table*}

%% file: TablesArxiv/Tab6.tex
\begin{table*}[t]
\caption{Size of $H_0: \sigma_t = \sigma$ for varying $\alpha$}
\setlength{\tabcolsep}{0.6cm}
\begin{center}
\footnotesize
\begin{tabular*}{0.99\columnwidth}{@{\extracolsep{\stretch{1}}}lr@{~~~}r@{~~~}r@{~~~}r@{~~~}r@{~~~}r@{~~~}r@{~~~}r@{~~~}r@{~~~}r@{~~~}r@{~}} 
 \multicolumn{12}{@{}l}{\it Panel A: $\alpha = -0.1250$} \\
\toprule
 & \multicolumn{3}{c}{$L^1$} & & \multicolumn{3}{c}{$L^2$} & & \multicolumn{3}{c}{$\sup$}  \\  
\cmidrule{2-12}
N & $0.01$ & $0.05$ & $0.10$ & & $0.01$ & $0.05$ & $0.10$ & & $0.01$ & $0.05$ & $0.10$ \\[0.22cm]
$     50 $ & $  0.005 $ & $  0.035 $ & $  0.078 $& & $  0.005 $ & $  0.036 $ & $  0.076 $& & $  0.002 $ & $  0.017 $ & $  0.040 $\\
$    100 $ & $  0.007 $ & $  0.041 $ & $  0.086 $& & $  0.007 $ & $  0.041 $ & $  0.084 $& & $  0.003 $ & $  0.026 $ & $  0.055 $\\
$    200 $ & $  0.008 $ & $  0.046 $ & $  0.096 $& & $  0.009 $ & $  0.045 $ & $  0.093 $& & $  0.006 $ & $  0.034 $ & $  0.069 $\\
$    500 $ & $  0.009 $ & $  0.048 $ & $  0.097 $& & $  0.009 $ & $  0.047 $ & $  0.098 $& & $  0.007 $ & $  0.040 $ & $  0.083 $\\
$   1000 $ & $  0.009 $ & $  0.047 $ & $  0.098 $& & $  0.009 $ & $  0.047 $ & $  0.096 $& & $  0.008 $ & $  0.040 $ & $  0.085 $\\
$   2000 $ & $  0.009 $ & $  0.049 $ & $  0.099 $& & $  0.010 $ & $  0.049 $ & $  0.098 $& & $  0.009 $ & $  0.043 $ & $  0.089 $\\
\bottomrule 
\end{tabular*}
\end{center}
\begin{center}
\footnotesize
\begin{tabular*}{0.99\columnwidth}{@{\extracolsep{\stretch{1}}}lr@{~~~}r@{~~~}r@{~~~}r@{~~~}r@{~~~}r@{~~~}r@{~~~}r@{~~~}r@{~~~}r@{~~~}r@{~}} 
 \multicolumn{12}{@{}l}{\it Panel B: $\alpha = 0$} \\
\toprule
 & \multicolumn{3}{c}{$L^1$} & & \multicolumn{3}{c}{$L^2$} & & \multicolumn{3}{c}{$\sup$}  \\  
\cmidrule{2-12}
N & $0.01$ & $0.05$ & $0.10$ & & $0.01$ & $0.05$ & $0.10$ & & $0.01$ & $0.05$ & $0.10$ \\[0.22cm]
$     50 $ & $  0.004 $ & $  0.030 $ & $  0.068 $& & $  0.004 $ & $  0.028 $ & $  0.067 $& & $  0.002 $ & $  0.015 $ & $  0.034 $\\
$    100 $ & $  0.007 $ & $  0.038 $ & $  0.083 $& & $  0.007 $ & $  0.037 $ & $  0.082 $& & $  0.004 $ & $  0.022 $ & $  0.052 $\\
$    200 $ & $  0.008 $ & $  0.044 $ & $  0.093 $& & $  0.008 $ & $  0.043 $ & $  0.091 $& & $  0.006 $ & $  0.032 $ & $  0.070 $\\
$    500 $ & $  0.010 $ & $  0.050 $ & $  0.098 $& & $  0.010 $ & $  0.051 $ & $  0.098 $& & $  0.008 $ & $  0.041 $ & $  0.084 $\\
$   1000 $ & $  0.010 $ & $  0.050 $ & $  0.101 $& & $  0.010 $ & $  0.049 $ & $  0.100 $& & $  0.009 $ & $  0.043 $ & $  0.087 $\\
$   2000 $ & $  0.010 $ & $  0.050 $ & $  0.100 $& & $  0.010 $ & $  0.050 $ & $  0.100 $& & $  0.009 $ & $  0.045 $ & $  0.094 $\\
\bottomrule 
\end{tabular*}
\end{center}
\begin{center}
\footnotesize
\begin{tabular*}{0.99\columnwidth}{@{\extracolsep{\stretch{1}}}lr@{~~~}r@{~~~}r@{~~~}r@{~~~}r@{~~~}r@{~~~}r@{~~~}r@{~~~}r@{~~~}r@{~~~}r@{~}} 
 \multicolumn{12}{@{}l}{\it Panel C: $\alpha = 0.1250$} \\
\toprule
 & \multicolumn{3}{c}{$L^1$} & & \multicolumn{3}{c}{$L^2$} & & \multicolumn{3}{c}{$\sup$}  \\  
\cmidrule{2-12}
N & $0.01$ & $0.05$ & $0.10$ & & $0.01$ & $0.05$ & $0.10$ & & $0.01$ & $0.05$ & $0.10$ \\[0.22cm]
$     50 $ & $  0.005 $ & $  0.029 $ & $  0.063 $& & $  0.005 $ & $  0.028 $ & $  0.062 $& & $  0.003 $ & $  0.015 $ & $  0.034 $\\
$    100 $ & $  0.006 $ & $  0.035 $ & $  0.074 $& & $  0.006 $ & $  0.036 $ & $  0.075 $& & $  0.004 $ & $  0.023 $ & $  0.048 $\\
$    200 $ & $  0.009 $ & $  0.042 $ & $  0.083 $& & $  0.009 $ & $  0.041 $ & $  0.085 $& & $  0.006 $ & $  0.031 $ & $  0.065 $\\
$    500 $ & $  0.009 $ & $  0.047 $ & $  0.096 $& & $  0.009 $ & $  0.048 $ & $  0.096 $& & $  0.007 $ & $  0.040 $ & $  0.083 $\\
$   1000 $ & $  0.011 $ & $  0.052 $ & $  0.104 $& & $  0.011 $ & $  0.053 $ & $  0.106 $& & $  0.010 $ & $  0.047 $ & $  0.094 $\\
$   2000 $ & $  0.012 $ & $  0.057 $ & $  0.108 $& & $  0.012 $ & $  0.057 $ & $  0.109 $& & $  0.011 $ & $  0.052 $ & $  0.102 $\\
\bottomrule 
\end{tabular*}
\end{center}
{\footnotesize \it  Size of the test $H_0: \sigma_t = \sigma$ for all $t$ with $\alpha = -0.1250$ (Panel A), $\alpha = 0$ (Panel B) and $\alpha = 0.1250$ (Panel C) and the three metrics $L^1, L^2$ and $\sup.$ Nominal sizes shown in top row are $1\%,$ $5\%$ and $10\%.$ $\lambda = 1$ and $20,000$ Monte Carlo simulations.}
\label{tab:RRVsizeTab}
\end{table*}

%% file: TablesArxiv/Tab7.tex
\begin{table*}[t]
\caption{Power of $H_0: \sigma_t =\sigma$ for $\alpha=-0.1250$ and varying $\beta$}
\setlength{\tabcolsep}{0.6cm}
\begin{center}
\footnotesize
\begin{tabular*}{0.99\columnwidth}{@{\extracolsep{\stretch{1}}}lr@{~~~}r@{~~~}r@{~~~}r@{~~~}r@{~~~}r@{~~~}r@{~~~}r@{~~~}r@{~~~}r@{~~~}r@{~}} 
\multicolumn{12}{@{}l}{\it Panel A: $\beta=0.05$} \\
\toprule
& \multicolumn{3}{c}{$L^1$} & & \multicolumn{3}{c}{$L^2$} & & \multicolumn{3}{c}{$\sup$} \\
\cmidrule(r){2-12}
N & 0.10 & 0.05 & 0.01 & & 0.10 & 0.05 & 0.01 & & 0.10 & 0.05 & 0.01 \\[0.22cm]  
     50  &  0.540 &  0.451 &  0.299 &&  0.545 &  0.460 &  0.308 &&  0.484 &  0.401 &  0.258\\
    100  &  0.747 &  0.670 &  0.533 &&  0.755 &  0.690 &  0.550 &&  0.729 &  0.663 &  0.533\\
    200  &  0.880 &  0.825 &  0.715 &&  0.886 &  0.842 &  0.741 &&  0.883 &  0.840 &  0.753\\
    500  &  0.965 &  0.944 &  0.887 &&  0.969 &  0.953 &  0.906 &&  0.970 &  0.958 &  0.922\\
   1000  &  0.989 &  0.979 &  0.949 &&  0.990 &  0.982 &  0.960 &&  0.991 &  0.987 &  0.973\\
   2000  &  0.998 &  0.996 &  0.987 &&  0.999 &  0.996 &  0.991 &&  0.999 &  0.996 &  0.993\\
\bottomrule
 \end{tabular*}
\end{center}

\begin{center}
\footnotesize
\begin{tabular*}{0.99\columnwidth}{@{\extracolsep{\stretch{1}}}lr@{~~~}r@{~~~}r@{~~~}r@{~~~}r@{~~~}r@{~~~}r@{~~~}r@{~~~}r@{~~~}r@{~~~}r@{~}} 
\multicolumn{12}{@{}l}{\it Panel B: $\beta=0.50$} \\
\toprule
& \multicolumn{3}{c}{$L^1$} & & \multicolumn{3}{c}{$L^2$} & & \multicolumn{3}{c}{$\sup$} \\
\cmidrule(r){2-12}
N & 0.10 & 0.05 & 0.01 & & 0.10 & 0.05 & 0.01 & & 0.10 & 0.05 & 0.01 \\[0.22cm]  
     50  &  0.523 &  0.423 &  0.274 &&  0.528 &  0.436 &  0.281 &&  0.460 &  0.380 &  0.231\\
    100  &  0.733 &  0.650 &  0.506 &&  0.748 &  0.664 &  0.535 &&  0.716 &  0.642 &  0.513\\
    200  &  0.864 &  0.811 &  0.702 &&  0.875 &  0.830 &  0.727 &&  0.876 &  0.831 &  0.741\\
    500  &  0.966 &  0.942 &  0.881 &&  0.972 &  0.953 &  0.901 &&  0.974 &  0.963 &  0.918\\
   1000  &  0.985 &  0.977 &  0.951& &  0.986 &  0.981 &  0.962 &&  0.989 &  0.986 &  0.970\\
   2000  &  0.998 &  0.990 &  0.982 &&  0.998 &  0.995 &  0.986 &&  0.999 &  0.998 &  0.990\\
\bottomrule
 \end{tabular*}
\end{center}

\begin{center}
\footnotesize
\begin{tabular*}{0.99\columnwidth}{@{\extracolsep{\stretch{1}}}lr@{~~~}r@{~~~}r@{~~~}r@{~~~}r@{~~~}r@{~~~}r@{~~~}r@{~~~}r@{~~~}r@{~~~}r@{~}} 
\multicolumn{12}{@{}l}{\it Panel C: $\beta=5.00$} \\
\toprule
& \multicolumn{3}{c}{$L^1$} & & \multicolumn{3}{c}{$L^2$} & & \multicolumn{3}{c}{$\sup$} \\
\cmidrule(r){2-12}
N & 0.10 & 0.05 & 0.01 & & 0.10 & 0.05 & 0.01 & & 0.10 & 0.05 & 0.01 \\[0.22cm]  
     50  &  0.341 &  0.234 &  0.121 &&  0.329 &  0.239 &  0.121 &&  0.272 &  0.190 &  0.096\\
    100  &  0.503 &  0.393 &  0.226 &&  0.524 &  0.417 &  0.259 &&  0.484 &  0.391 &  0.254\\
    200  &  0.706 &  0.606 &  0.435 &&  0.729 &  0.626 &  0.469 &&  0.715 &  0.635 &  0.483\\
    500  &  0.898 &  0.828 &  0.693 &&  0.905 &  0.850 &  0.733 &&  0.915 &  0.870 &  0.774\\
   1000  &  0.965 &  0.935 &  0.855 &&  0.972 &  0.946 &  0.882 &&  0.982 &  0.960 &  0.915\\
   2000  &  0.993 &  0.980 &  0.941 &&  0.994 &  0.987 &  0.954 &&  0.997 &  0.992 &  0.976\\
\bottomrule
 \end{tabular*}
\end{center}
{\footnotesize \it Power of the test $H_0: \sigma_t = \sigma$ for all $t$ for $\alpha=-0.1250,$ $\beta = 0.05$ (Panel A), $\beta = 0.5$ (Panel B) and $\beta = 5$ (Panel C), where we simulate the stochastic volatility process under the alternative hypothesis as $\log \sigma^2_t = \int_{-\infty}^t e^{-\beta(t-s)}dB_s,$ see Section \ref{sec:simset}. $\lambda=1$ and we consider the three metrics $L^1, L^2$ and $\sup.$ $20,000$ Monte Carlo simulations.}
\label{tab:RRVpowPlus}
\end{table*}

%% file: TablesArxiv/Tab8.tex
\begin{table*}[t]
\caption{Power of $H_0: \sigma_t =\sigma$ for $\alpha=0.1250$ and varying $\beta$}
\setlength{\tabcolsep}{0.6cm}
\begin{center}
\footnotesize
\begin{tabular*}{0.99\columnwidth}{@{\extracolsep{\stretch{1}}}lr@{~~~}r@{~~~}r@{~~~}r@{~~~}r@{~~~}r@{~~~}r@{~~~}r@{~~~}r@{~~~}r@{~~~}r@{~}} 
\multicolumn{12}{@{}l}{\it  Panel A: $\beta=0.05$}\\
\toprule
& \multicolumn{3}{c}{$L^1$} & & \multicolumn{3}{c}{$L^2$} & & \multicolumn{3}{c}{$\sup$} \\
\cmidrule(r){2-12}
N & 0.10 & 0.05 & 0.01 & & 0.10 & 0.05 & 0.01 & & 0.10 & 0.05 & 0.01 \\[0.22cm]  
     50  &  0.495 &  0.417 &  0.265 &&  0.499 &  0.428 &  0.285 &&  0.441 &  0.364 &  0.234\\
    100  &  0.699 &  0.625 &  0.493 &&  0.712 &  0.645 &  0.516 &&  0.676 &  0.618 &  0.491\\
    200  &  0.850 &  0.802 &  0.688 &&  0.857 &  0.817 &  0.715 &&  0.855 &  0.812 &  0.725\\
    500  &  0.964 &  0.944 &  0.882 &&  0.968 &  0.954 &  0.904 &&  0.968 &  0.956 &  0.919\\
   1000  &  0.988 &  0.979 &  0.950 &&  0.990 &  0.980 &  0.960 &&  0.992 &  0.987 &  0.972\\
   2000  &  0.998 &  0.996 &  0.986 &&  0.999 &  0.997 &  0.990 &&  0.999 &  0.998 &  0.992\\
\bottomrule
 \end{tabular*}
\end{center}

\begin{center}
\footnotesize
\begin{tabular*}{0.99\columnwidth}{@{\extracolsep{\stretch{1}}}lr@{~~~}r@{~~~}r@{~~~}r@{~~~}r@{~~~}r@{~~~}r@{~~~}r@{~~~}r@{~~~}r@{~~~}r@{~}} 
\multicolumn{12}{@{}l}{\it Panel B: $\beta=0.50$}\\
\toprule
& \multicolumn{3}{c}{$L^1$} & & \multicolumn{3}{c}{$L^2$} & & \multicolumn{3}{c}{$\sup$} \\
\cmidrule(r){2-12}
N & 0.10 & 0.05 & 0.01 & & 0.10 & 0.05 & 0.01 & & 0.10 & 0.05 & 0.01 \\[0.22cm]  
     50  &  0.470 &  0.384 &  0.251 &&  0.472 &  0.394 &  0.253 &&  0.421 &  0.339 &  0.207\\
    100  &  0.682 &  0.611 &  0.470 &&  0.703 &  0.627 &  0.488 &&  0.666 &  0.598 &  0.469\\
    200  &  0.838 &  0.787 &  0.672 &&  0.855 &  0.803 &  0.690 &&  0.845 &  0.804 &  0.702\\
    500  &  0.967 &  0.944 &  0.876 &&  0.971 &  0.951 &  0.896 &&  0.970 &  0.959 &  0.919\\
   1000  &  0.986 &  0.978 &  0.951 &&  0.988 &  0.981 &  0.959 &&  0.991 &  0.986 &  0.973\\
   2000  &  0.998 &  0.991 &  0.981 &&  0.999 &  0.995 &  0.985 &&  1.000 &  0.999 &  0.992\\
\bottomrule
 \end{tabular*}
\end{center}
\begin{center}
\footnotesize
\begin{tabular*}{0.99\columnwidth}{@{\extracolsep{\stretch{1}}}lr@{~~~}r@{~~~}r@{~~~}r@{~~~}r@{~~~}r@{~~~}r@{~~~}r@{~~~}r@{~~~}r@{~~~}r@{~}} 
\multicolumn{12}{@{}l}{\it Panel C: $\beta=5.00$}\\
\toprule
& \multicolumn{3}{c}{$L^1$} & & \multicolumn{3}{c}{$L^2$} & & \multicolumn{3}{c}{$\sup$} \\
\cmidrule(r){2-12}
N & 0.10 & 0.05 & 0.01 & & 0.10 & 0.05 & 0.01 & & 0.10 & 0.05 & 0.01 \\[0.22cm]  
     50  &  0.297 &  0.209 &  0.104 &&  0.288 &  0.216 &  0.113 &&  0.232 &  0.172 &  0.086\\
    100  &  0.478 &  0.377 &  0.208 &&  0.496 &  0.396 &  0.242 &&  0.454 &  0.373 &  0.235\\
    200  &  0.687 &  0.590 &  0.413 &&  0.703 &  0.617 &  0.445 &&  0.688 &  0.610 &  0.456\\
    500  &  0.901 &  0.836 &  0.690 &&  0.907 &  0.859 &  0.731 &&  0.918 &  0.875 &  0.769\\
   1000  &  0.966 &  0.931 &  0.851 &&  0.972 &  0.944 &  0.879 &&  0.977 &  0.960 &  0.915\\
   2000  &  0.992 &  0.981 &  0.943 &&  0.994 &  0.988 &  0.954 &&  0.996 &  0.992 &  0.976\\
\bottomrule
 \end{tabular*}
\end{center}
{\footnotesize \it Power of the test $H_0: \sigma_t = \sigma$ for all $t$ for $\alpha=0.1250,$ $\beta = 0.05$ (Panel A), $\beta = 0.5$ (Panel B) and $\beta = 5$ (Panel C), where we simulate the stochastic volatility process under the alternative hypothesis as $\log \sigma^2_t = \int_{-\infty}^t e^{-\beta(t-s)}dB_s,$ see Section \ref{sec:simset}. $\lambda=1$ and we consider the three metrics $L^1, L^2$ and $\sup.$ $20,000$ Monte Carlo simulations.}
\label{tab:a6betapow}
\end{table*}